\documentclass[12pt]{article}
\usepackage[toc,page]{appendix}
\usepackage{jheppub}
\usepackage{setspace}
\usepackage{amssymb}
\usepackage{amsmath,amsfonts}
\usepackage{graphicx}
\usepackage{mathrsfs}
\usepackage[vcentermath]{youngtab}
\usepackage{color}
\usepackage{tikz}
\usetikzlibrary{shapes,arrows,shadows}
\numberwithin{equation}{section}
\usepackage{bbm}

\def\bea{\begin{eqnarray}}
\def\eea{\end{eqnarray}}

\newcommand{\eq}{\begin{equation}}
\newcommand{\en}{\end{equation}}
\newcommand{\eqn}{\begin{eqnarray}}
\newcommand{\enn}{\end{eqnarray}}
\newcommand{\nn}{\nonumber }
\newcommand{\beq}{\begin{equation}}
\newcommand{\eeq}{\end{equation}}

\newcommand{\cX}{\mathcal{X}}
\newcommand{\cY}{\mathcal{Y}}

\newcommand{\cN}{\mathcal{N}}

\newcommand{\gd}[1][{}]{\delta_{#1}{}}

\usepackage[normalem]{ulem}

\makeatletter
\gdef\@fpheader{}
\makeatother

\begin{document}

\title{ EXCEPTIONALITY , SUPERSYMMETRY AND NON-ASSOCIATIVITY IN PHYSICS }
\author{ Murat G\"unaydin }

\affiliation{ 
Institute for Gravitation and the Cosmos \\
Physics Department, \\
Pennsylvania State University \\
University Park, PA 16802, USA}

\emailAdd{mgunaydin@psu.edu}
\abstract{This is an expanded and updated version of the talk I gave at the meeting organized at CERN on April 27-28, 2015 dedicated to honoring the memory of Bruno Zumino and his legacy. In this talk I  review the emergence of exceptional structures and non-associative algebras in quantum mechanics, supergravity and M/superstring theories.  }
\maketitle

\section{Introduction}

It is a great honor for me to give a talk at a meeting dedicated to Bruno Zumino's memory and his scientific legacy. His papers with Julius Wess on spacetime supersymmetry and the developments they led to had a tremendous impact on my career.  
I first met Bruno Zumino during the summer of 1977 during a short visit to CERN. Later I had the privilege of interacting with him  during my year  at CERN in (1981-82) and  at Lawrence Berkeley Lab during the period 1984-86. I co-authored two papers with Bruno Zumino. Our first paper was joint work  with John Ellis and Mary K. Gaillard  titled  "Supersymmetry and Non-compact groups in Supergravity" and published in Nuclear Physics  B \cite{Ellis:1983na}.
  My second paper with Bruno Zumino was written at Lawrence Berkeley Lab. titled "Magnetic Charge and Non-Associative Algebras", that appeared in the proceedings of a meeting in honor of G.C. Wick held at Scuola Normale Superiore in Pisa \cite{ Gunaydin:1985ur}.

 During our first meeting in 1977 Bruno Zumino wanted to talk to me about octonions. He knew about my work with Feza G\"ursey on color quarks and octonions and my first paper on space-time supersymmetry which also involved octonions.  Later in Lawrence Berkeley Lab we worked  on the question of how non-associativity arises in the presence of  magnetic charges and  how non-associativity might  be integrated into the framework of quantum mechanics. Two years  before he passed away I wrote a paper with D. Minic, which we dedicated to Bruno Zumino on the occasion of his 90th birthday, in which we pointed out that a non-associative algebra that had appeared recently in closed string theory is isomorphic  to the  non-associative  magnetic algebra Bruno Zumino and I studied in our paper. 
 
 Bruno Zumino's  last three papers are on supergravity theories defined by Jordan algebras and associated Freudenthal triple systems, in particular the exceptional Jordan algebra and its Freudenthal triple system\cite{Cerchiai:2010xv,Ferrara:2012cr,Marrani:2012uu}.  Since all the exceptional groups are related to octonions and exceptional Jordan algebra and they  arise in some remarkable supergravity theories   I decided to talk about "Exceptionality, supersymmetry and non-associativity in Physics" in this meeting dedicated to his memory. 
 
  Broad outline of the topics I decided to talk about were:
 \begin{itemize}
 \item Jordan formulation of quantum mechanics.
 \item Octonionic quantum mechanics and the exceptional Jordan algebra.
 \item Connections between exceptional Lie algebras and the exceptional Jordan algebra. Magic square of Freudenthal, Rozenfeld and Tits.
\item Appearance of exceptional  groups as global symmetry groups in supergravity.
 \item Orbits of extremal black hole solutions under the action of their U-duality groups and Jordan algebras and Freudenthal triple systems
 \item Non-associative "magnetic algebra" of G\"unaydin and Zumino and its appearance in closed string theory. \item Generalization of the magnetic algebra and Stuckelberg's generalization of classical Poisson brackets that do not satisfy the Jacobi identity.
\item Superspaces defined by Jordan superalgebras and the exceptional superspace that has no realization in terms of associative (super)matrices.
\item Super analogs of the Magic Square and existence of novel  simple superalgebras over finite fields that have no counterpart in characteristic zero. Fascinating surprises!
\end{itemize}  

However since the writing up my talk has been in gestation for a very long time I will include some of the results obtained since my talk that are directly related to the above topics in this write-up.   In particular I will summarize the results of my work with L\"ust and Malek in section \ref{GLM} related to my paper  with Bruno Zumino on magnetic charges and Malcev algebras. 

\section{ Jordan  Formulation of Quantum Mechanics}

In the years 1932-33 Pascual Jordan proposed a novel formulation of  quantum mechanics (QM)  that came to be called Jordan formulation\cite{Jordan:1933}. His main motivation was to generalize the quantum mechanical formalism so as to be able to accommodate the then observed beta decay phenomena  within this generalized framework!
Since the  commutator  of two  Hermitian operators corresponding  to observables that act on an Hilbert space is not Hermitian:
\eq H_1^\dagger =H_1 \quad , \quad  H_2^\dagger =H_2 \longrightarrow  \quad [H_1,H_2]^\dagger \neq [H_1,H_2] \en
he proposed using the symmetric anti-commutator product among Hermitian operators under which they remain Hermitian. Under the symmetric product  $ H_1\circ H_2 = 1/2 ( H_1 H_2 + H_2 H_1) $  Hermitian operators $H_1$ and $H_2$ acting on the physical Hilbert space satisfy the identities:
\eqn H_1 \circ H_2 = H_2 \circ H_1= (H_1 \circ H_2)^\dagger \\ 
H_1\circ ( H_2 \circ H_1^2 ) = (H_1 \circ H_2 ) \circ H_1^2 \enn
which are taken as defining identities of a Jordan algebra. Jordan hoped that there would be a rich family of algebras satisfying the above identities and which can not be realized in terms of linear operators acting on a vector ( Hilbert) space with the product taken as 1/2 their anticommutator.

In the formulation of quantum mechanics over an Hilbert space a la Dirac the pure states are represented by rays over the Hilbert space and in the Jordan formulation these states are represented by irreducible projection operators ( idempotents). Furthermore, these irreducible projection operators correspond to propositions ( yes-no experiments) in the propositional calculus approach of Birkhoff and von Neumann who showed that the axioms of quantum mechanics are equivalent to the axioms of projective geometry \cite{MR1435976}. Hence we can represent the connections between different formulations of QM and projective geometry with the following chart
\newpage
\begin{center}
 {\bf Dirac formulation of quantum mechanics \\ over an Hilbert space }\\
$\quad \quad $  Pure states  $\, \,  \quad  \quad \quad \quad  \quad   \Longleftrightarrow \quad \quad \, \quad  \,$ rays $|\psi \rangle q $ with $\bar{q} q =1$ \\
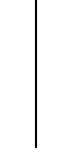
\begin{tikzpicture}
\draw[thick] (4,4) -- (4,6);
\end{tikzpicture} 

{\bf Jordan formulation of quantum mechanics } \\
 Pure states $ \quad \quad \quad \quad  \quad \,\, \Longleftrightarrow  \,\,$  $ \quad \quad  | \psi \rangle \langle \psi |= P_\psi = P^2_\psi$ \\
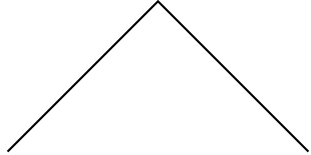
\begin{tikzpicture}
\draw[thick] (2,2) --+ (2,2) --+ (4,0);
\end{tikzpicture}
\end{center}
\begin{displaymath}
  \begin{array}{ccc}
 \rm {\bf Propositional \, \, \, calculus } &\quad \quad \, \,\, \Longleftrightarrow & \rm {\bf Projective \, \,  geometry } \\
  \rm {\bf  Axioms \, of \, quantum \, mechanics } & \quad  \quad \quad \Longleftrightarrow & \rm {\bf Axioms \, of \,  projective \, geometry }\\
\rm {\bf Pure \,  states: \,  Tr(P_\psi )=1 } & \quad \quad \quad  \Longleftrightarrow  & \rm {\bf Points \, in \, projective \, geometry }\\
\rm {\bf Superpositions \, of \,  | \psi \rangle \, and \, |\xi \rangle } & \quad \quad \quad  \Longleftrightarrow & \rm  \quad {\bf Line  \,  connecting \, points \, \,    P_\psi \,  and \,  P_\xi }
 \end{array}
\end{displaymath}
Jordan , von Neumann and Wigner (JvNW) (1934) gave a complete classification of all finite dimensional simple  Jordan algebras and showed that with one exception all finite dimensional simple Jordan algebras are  special  i.e. they can be realized as linear operators acting on a vector space or Hilbert space with the product being 1/2 the anticommutator \cite{Jordan:1933vh}.
  The complete list of finite dimensional simple Euclidean Jordan algebras are as follows\footnote{A Jordan algebra $J$\  is called Euclidean if $X^2 +Y^2=0 $ implies that $X=Y=0$ for any two elements of $J$.}: \\
i) Dirac gamma matrices $\Gamma(d)$ in $d$   Euclidean dimensions under the symmetric anticommutator product. \\
ii) Jordan algebras  $J_n^{\mathbb{R}} \quad J_n^{\mathbb{C}} \quad J_n^{\mathbb{H}} $
generated by  $n\times n$  Hermitian  matrices over reals $\mathbb{R}$, complex numbers $\mathbb{C}$ and quaternions $\mathbb{H}$ 
with the Jordan product being 1/2 the anticommutator:
\[ A\circ B \equiv \frac{1}{2} ( AB +BA) \]
iii) {\color{black} Exceptional }Jordan algebra $J_3^{\mathbb{O}}$ of $3\times 3$ Hermitian matrices over the division algebra of octonions $\mathbb{O}$, which  can not be realized in terms of linear operators acting on some vector space.

Hence the proposal of Pascual Jordan to generalize the algebraic  framework of quantum mechanics led to a single novel algebraic structure, namely the exceptional Jordan algebra $J_3^{\mathbb{O}}$, in the finite dimensional case.   Let alone explaining the beta decay phenomena  it was not at all obvious at the time  if all the axioms of quantum mechanics could be implemented in the exceptional case due to its intrinsic non-associativity.

    \section{ Octonionic Quantum Mechanics  and the quadratic Jordan formulation of Quantum Mechanics}
 Notwitstanding the intrinsic non-associativity of the exceptional Jordan algebra $J_3^{\mathbb{O}}$ it was shown by G\"unaydin, Piron and Ruegg (GPR) that one can formulate quantum mechanics over $J_3^{\mathbb{O}}$ satisfying all the axioms of QM as formulated by  von Neumann\cite{Gunaydin:1978jq}. The pure states in the corresponding QM are represented by
irreducible  projection operators ( idempotents)  belonging to the
exceptional Jordan algebra $J_3^{\mathbb{O}}$ over the division algebra of  real octonions $\mathbb{O}$. Corresponding projective geometry is  the
octonionic Moufang plane,
 which can be orthocomplemented with  a unique probability function.
The result of successive,
compatible experiments do not depend  on the order in which they are
performed, in spite of the non-associativity of underlying octonions. The  QM described by the exceptional Jordan algebra is referred to as octonionic QM and has no known Hilbert Space formulation.
  The quantum mechanics described by the exceptional Jordan algebra $J_3^\mathbb{O}$ describes two octonionic degrees of freedom and its projective geometry is non-Desarguesian. We can represent the connections between different aspects of octonionic quantum mechanics with the following chart:
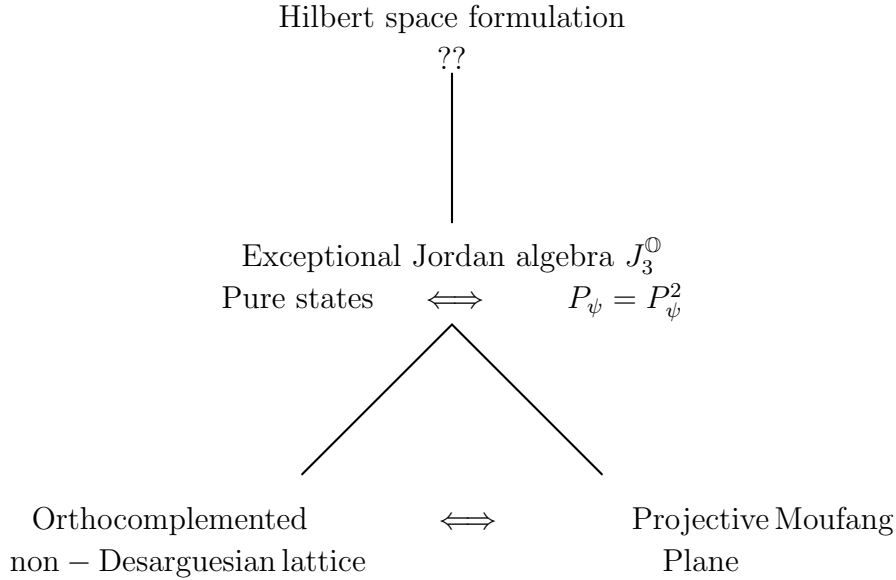
\begin{center}
 Hilbert space formulation\\
?? \\
\begin{tikzpicture}
\draw[thick] (4,4) -- (4,6);
\end{tikzpicture}

Exceptional Jordan algebra $J_3^{\mathbb{O}}$ \\
 Pure states $  \quad \,\, 
 \Longleftrightarrow  \,\,$  $ \quad \quad  P_\psi = P^2_\psi$ \\
\begin{tikzpicture}
\draw[thick] (2,2) --+ (2,2) --+ (4,0);
\end{tikzpicture}
\end{center}
\begin{displaymath}
  \begin{array}{ccc}
 \rm Orthocomplemented  \, \quad  & \quad \quad \Longleftrightarrow & \quad \qquad \quad \rm Projective \, Moufang  \\
\rm  non-Desarguesian  \, lattice & & \rm Plane \\
 \end{array}
\end{displaymath}

Later the Jordan formulation of quantum mechanics  in terms of the symmetric Jordan product was extended to the quadratic Jordan formulation in terms of the Jordan triple product \cite{Gunaydin:1979df,Gunaydin:1989ur,Gunaydin:1989mgm}. The Jordan triple product is defined as   \eq \{ABC\} \equiv  ( A\circ B) \circ C + A \circ ( B \circ C) -(A\circ C) \circ B \en
where $ A\circ B = \frac{1}{2} ( AB+BA)  $. \\
 Quadratic Jordan formulation extends also to the octonionic quantum mechanics in a straightforward manner. In the quadratic formulation  the commutator of two Hermitian operators  corresponding to observables goes over to the commutator of the quadratic actions of these two Hermitian operators which involve only the Jordan triple product as indicated in the Table \ref{quadratic} .  Furthermore the quadratic Jordan  formulation of quantum mechanics can be extended formally to formulation of quantum mechanics over finite fields as well as to Jordan superalgebras including the exceptional Jordan superalgebra to be discussed later  \cite{Gunaydin:1979df,Gunaydin:1989mgm}. 
 However, one loses , in general, the probability interpretation since the corresponding trace function is  no longer a real number.

\begin{table}
 \begin{displaymath}
\begin{array}{|c|c|c|}
\hline
~&~&~\\
 Hilbert \, Space &  Jordan &  Quadratic \, Jordan \\
 Formulation &  Formulation &  Formulation \\
\hline
~&~&~\\
\mid \alpha > &  \mid \alpha ><\alpha \mid~=~P_{\alpha} &
P_{\alpha}\\
~&~&~\\
H \mid \alpha > & H \circ P_{\alpha} & \Pi_{H}P_{\alpha} =
\{HP_{\alpha}H\} \\
~&~&~\\
<\alpha \mid H \mid \beta > & ? & ?\\
~&~&~\\
<\alpha \mid H \mid \alpha > & Tr~H \circ P_{\alpha} &
Tr~\Pi_{P_{\alpha}}H ~=~ Tr~\{P_{\alpha} HP_{\alpha}\}\\
~&~&~\\
\mid<\alpha \mid H \mid \beta >\mid^{2} & Tr~P_{\alpha} \circ
\{HP_{\beta}H\} & Tr~ \Pi_{P_{\alpha}}\Pi_{H}P_{\beta}\\
~&~&~\\
~&=~ Tr~\{HP_{\alpha}H\} \circ P_{\beta} & =~
 Tr~ \Pi_{P_{\beta}}\Pi_{H}P_{\alpha}\\
~&~&~\\
~[H_{1},H_{2}] & ? & 4(\Pi_{H_{1}} \circ \Pi_{H_{2}} -
\Pi_{H_{1} \circ H_{2}})\\
~&~&~   \\
\hline
\end{array}
\end{displaymath}
\caption{ \label{quadratic} Comparison of Dirac formulation, Jordan formulation and quadratic Jordan formulation of quantum mechanics. $\{  .,.,.\} $ denotes the Jordan triple product and $\Pi_H$ is the quadratic operator defined as $\Pi_H X \equiv \{H,X,H\}$. $Tr$ denotes the trace function.   }
\end{table}

With the inclusion of the space-time degrees of freedom any finite dimensional quantum mechanics becomes infinite dimensional. This suggests that  there must exist some infinite dimensional exceptional Jordan algebras that describe  extensions of the finite dimensional octonionic quantum mechanics  when spacetime degrees are incorporated. However in a remarkable series of papers Zelmanov  showed that there are no  infinite dimensional exceptional Jordan algebras \cite{MR688595,MR566787, MR566779}. Zelmanov's results on Jordan algebras during the period 1979-1983  were referred to as "Russian revolution in Jordan algebras" by McCrimmon\cite{MR744730}. These results imply that to be able to extend octonionic quantum mechanics in a higher dimensional quantum mechanics one has to generalize or modify the axioms of quantum mechanics so that the underlying geometry is no longer projective since a non-Desarguesian projective geometry can not be embedded in a higher projective one. In particular this implies going beyond Jordan algebras as the relevant underlying algebraic structures.\footnote{
Indeed the algebra of $n\times n$ Hermitian matrices over octonions under the anti-commutator product   is no longer a Jordan algebra for $ n>3$.}

\section{Exceptional Jordan Algebra and exceptional Groups}
 Prior to the discovery of the exceptional Jordan algebra by Jordan, von Neumann and Wigner (JvNW)  it was known that the smallest exceptional group $G_2$ is the automorphism group of octonions.
 After the work of JvNW mathematicians established deep connections between the exceptional Jordan algebra and the other exceptional Lie groups $F_4, E_6 , E_7 ,E_8$.  We refer the reader to the famous review by Freudenthal\cite{MR0170974} about the connections between exceptional Lie groups, the exceptional Jordan algebra and the remarkable geometries whose symmetry groups form the famous Magic Square. 
 
 There is a beautiful unified  construction of the Lie algebras of the Magic Square    due to Tits\cite{MR0146231,Tits66}.\footnote{The review paper of Freudenthal as well as the relevant papers of Tits were first brought to my attention  by Nathan Jacobson during the Spring of 1971 and I had the good fortune to attend his course on exceptional Lie algebras  during the Fall of 1971 at Yale. 
 I am deeply grateful to him and  to George Seligman for many enlightening discussions on exceptional Lie algebras and related algebraic structures during my graduate years at Yale and for very helpful correspondence in the subsequent years.}    
\begin{table}
\begin{displaymath}
\begin{array}{|c|c|c|c|c|}
\hline
~ & J_3^{\mathbb{R}} &J_3^{\mathbb{C}} &J_3^{\mathbb{H}}&J_3^{\mathbb{O}} \\ \hline
&&&& \\
~ \mathbb{R} ~ & SO(3) & SU(3) & ~ USp(6) ~ & ~~ ~F_4 ~~~\\
\hline
&&&& \\
 \mathbb{C} & SU(3) & SU(3)\times SU(3)  & SU(6) & E_6 \\
 \hline
 &&&& \\
  \mathbb{H} &~ USp(6) ~ & SU(6)  & SO(12) & E_7 \\
 \hline
 &&&&     \\
  \mathbb{O} &F_4& E_6  & E_7 & E_8 \\
 \hline
\end{array}
\end{displaymath}
\caption{ \label{magicsquare} Magic Square of Freudenthal, Rozenfeld and Tits}
\end{table}
The Tits construction of the Lie algebras of the Magic Square is given over the tensor product of composition algebras with the Jordan algebras of degree three over them. Lie algebra in Tits' construction  can be given a vector  space decomposition of the form:
\eq L= Der(\mathbb{A}) \oplus \mathbb{A}_0 \otimes J_{3_0}^{\mathbb{A'}} \oplus Der (J_{3}^{\mathbb{A'}}) \en
where $\mathbb{A}$ and $\mathbb{A'}$ are composition algebras and $J_{3}^{\mathbb{A'}}$ denotes a Jordan algebra of degree three defined by $3\times3 $ Hermitian matrices over $\mathbb{A'}$.  $Der (\mathbb{A}) $ denotes the derivation algebra of the composition algebra $\mathbb{A}$ and $Der(J_{3}^{\mathbb{A'}})$ denotes the derivation algebra  of $J_{3}^{\mathbb{A'}}$, respectively\footnote{ Derivation $D$ of an algebra $A$\ is a linear transformation that satisfies the Leibniz rule $D(XY)=(DX)Y+X(DY)$ and derivations form a Lie algebra under commutation, which is simply the Lie algebra of the inner automorphism group of $A$.}.  $\mathbb{A}_0$ denotes the subspace of pure imaginary elements of  $\mathbb{A}$ and  and $J_{3_0}^{\mathbb{A'}} $ denotes the subspace of traceless elements of $ J_{3}^{\mathbb{A'}}$.
The product of two general elements of $\mathbb{A}$ can be written as
\eq
ab=a*b + (a,b)
\en
where $(a,b)$ is an element of the underlying field which we take to be real numbers and $a*b$ is pure imaginary. Similarly one can decompose the product of two general elements $A,B$ of  $ J_{3}^{\mathbb{A'}}$ in the form
\eq
A \circ B= A*B + (A,B)\, \mathrm{I}_3
\en
where $A*B$ is a traceless element of  $ J_{3}^{\mathbb{A'}}$ and $(A,B)$ is the bilinear form  given by the trace  and $\mathrm{I}_3$ denotes the identity element of the Jordan algebra.

A general element of $Der(\mathbb{A})$ can be written as a linear operator $d_{a,b}$  labelled by a pair of elements $a,b$ of $\mathbb{A}$ whose action on a general element $x$ of $\mathbb{A}$ is given by:
\eq
d_{a,b}\, x = \frac{1}{4} [[a,b],x]-\frac{3}{4} [a,b,x]
\en
where $[a,b]=ab -ba $ and $[a,b,x]=(ab)x-a(bx)$ is the associator.  Derivations of $J_3^{\mathbb{A}'}$ are also labelled by a pair of elements $A,B$ of  $J_3^{\mathbb{A}'}$ whose action on a general element $X$ is defined as:
\eq
D_{A,B} \, X = A \circ (B\circ X)-B\circ (A\circ X)
\en
where $\circ$ denotes the Jordan product.  The commutation relations of the derivations with other elements of $L$ are defined as
\bea
[d+D,x\otimes X]&=& d(x)\otimes X+x\otimes D(X)  \\
{[} a\otimes A, b \otimes B] &=&(A,B)\, d_{a,b} + (a*b)\otimes (A*B)+ (a,b)\, D_{A,B}
\eea
where $a,b \in\mathbb{A}_0$ and $A,B \in J_{3_0}^{\mathbb{A}'}$ and
\eq
[d_{a,b},D_{A,B}]=0\en
The compact forms of the Lie algebras of the Magic Square obtained by Tits construction is given in Table \ref{magicsquare}. The fact that the resulting square of Lie algebras turned out to be symmetric was one of the reasons it was called "Magic Square".  Depending on whether one considers real or split composition algebras $\mathbb{A}$ and $\mathbb{A}'$ and compact or noncompact Jordan algebras one obtains different real forms of the Magic Square which were all determined by Tits\cite{MR0146231,Tits66,Titsnotes}.

\section{Octonions, color quarks and exceptional grand unified theories }

The compact forms of the groups in the  magic square contain basically all the internal symmetry groups of hadronic world that were studied by the physicists in the 1950s and 1960s except for the exceptional groups. It begged the question whether the exceptional groups could also be relevant for the physics  of elementary particles. This was part of the motivation of my work with G\"ursey on the possible role exceptional groups may play in Physics\cite{Gunaydin_thesis,Gunaydin:1973rs,Gunaydin:1974fb,Gunaydin:1976vq}. Furthermore, since the exceptional groups are all related to octonions we investigated the question whether
the observed internal symmetries of particle physics can be understood at a fundamental level by extending the underlying field of quantum mechanics from complex numbers to the division algebra of octonions. 
 Our  work led to the so-called algebraic confinement scheme in which quarks are represented by transverse octonionic fields with $SU(3)$ subgroup of the automorphism group $G_2$ of octonions is  identified as the color $SU(3)_C$. The states in color singlet sector are described by  an ordinary complex Hilbert  space. This proposal provided a  mathematical model of the original suggestion of Gell-Mann that color quarks operate in a {\bf fictitious} Hilbert space and that only the color  singlet sector corresponds to observables \cite{GellMann:1981ph}.
However our scheme did not incorporate any dynamics. Shortly afterwards quantum chromodynamics was established as the true dynamical theory of strong interactions in which the  quarks are {\bf real} but {\bf  confined}.

Subsequently  grand unified theories (GUT) based on exceptional groups were formulated.  
$E_6$ GUT was proposed by  G\"ursey, Ramond \& Sikivie\cite{Gursey:1975ki}, $E_7$ GUT was proposed by G\"ursey and Sikivie\cite{Gursey:1976dn} and by Ramond\cite{Ramond:1976jg}  and a family unifying GUT based on the largest exceptional group 
$E_8$ was proposed by Bars and G\"unaydin \cite{Bars:1980mb}. 
Later $E_8 \times E_8$ appeared as gauge symmetry of one of the heterotic string theories constructed by Gross, Harvey, Martinec and Rohm \cite{Gross:1984dd}.

\section{ Space-time  superalgebras, Jordan algebras and exceptional groups}

 I was an Accademia dei Lincei fellow at Scuola Normale Superiore in Pisa when I received the preprint of the famous paper of Bruno Zumino with Julius Wess on spacetime supersymmetry\cite{Wess:1974tw} which was inspired by the work on world-sheet supersymmetry in string theory.  I was immediately attracted to the idea of space-time supersymmetry and started working on it.  
 
 In the early days  of space-time supersymmetry 
two of the most important open  problems were the following:  \\
1. How to formulate a local gauge theory of spacetime supersymmetry that necessarily involves gravity, which came to be called supergravity. Simple supergravity was first formulated in \cite{Freedman:1976xh} and shortly thereafter also in \cite{Deser:1976eh}.\\
2. Is there an exceptional superalgebra whose local gauge theory would lead to a unified theory of all interactions including gravity and the standard model of particle physics with a single graviton? Here the term exceptional refers to a Lie superalgebra beyond the standard $N=8$ Poincare superalgebra whose massless supermultiplet with a single graviton has 28  vector fields transforming in the adjoint of  $SO(8)$  symmetry group which is not large enough to accommodate the standard model gauge group $SU(3)\times SU(2)\times U(1)$.   This point was especially stressed by Murray Gell-Mann. \\

 My first paper on  supersymmetry  was an attempt to try to answer the second question in which the  concept of generalized spacetimes coordinatized by Jordan algebras  was introduced\cite{Gunaydin:1975mp}. The basic idea for this proposal was based on the fact that
 Minkowski spacetime in four dimensions can be coordinatized by Hermitian $2\times 2$ matrices:
\[ x = \sigma_\mu x^\mu \] , where $\sigma_\nu= ( 1_2, \vec{\sigma} )$ are the Pauli matrices, can be considered as elements of the Jordan algebra $J_2^{\mathbb{C}}$ of $2\times 2$ Hermitian matrices over the field of  complex numbers $\mathbb{C}$ with the Jordan product taken as 1/2 the anticommutator.
 Considered as a Jordan algebra one then has the identifications:
 \begin{itemize}
 \item
 Automorphism group of $J_2^{\mathbb{C}} = SU(2) \Longrightarrow $  rotation group $Rot( J_2^{\mathbb{C}})$ of $J_2^{\mathbb{C}}$  \item
 Reduced structure group defined as the invariance group of the norm  $N(x)$ of $x$ given by $  \eta_{\mu\nu} x^\mu x^\nu = Det (x) =SL(2,C) \Longrightarrow $ Lorentz group $Lor( J_2^{\mathbb{C}}) $ of  $J_2^{\mathbb{C}}$.\item
 Linear fractional group (M\"obius)  of $J_2^{\mathbb{C}} = SU(2,2) \Longrightarrow $  Conformal group $ Conf( J_2^{\mathbb{C}})$ of  $J_2^{\mathbb{C}}$. \end{itemize}
In fact the coordinatization of Minkowski space-time in terms of $2\times 2$ matrices underlies the twistor theory of Penrose. 

 In my paper \cite{Gunaydin:1975mp} I listed the compact forms of the "spacetime symmetry" groups associated with simple Jordan algebras of degree three and stressing that depending on the real forms of the Jordan algebra one would obtain different real forms of these groups which were all known at the time \cite{MR0146231,Titsnotes}. In Table \ref{euclidean}  I list the symmetry groups of the simple Euclidean Jordan algebras of degree three over the four division algebras $\mathbb{R},\mathbb{C},\mathbb{H}$ and $\mathbb{O}$. In Table \ref{split}  I list the corresponding symmetry groups for simple Jordan algebras of degree three over the split composition algebras $\mathbb{R},\mathbb{C}_s,\mathbb{H}_s$ and $\mathbb{O}_s$.

My main goal was to decompose the components of the exceptional Jordan algebra over split octonions with respect to the four dimensional Lorentz subgroup $SL(2,\mathbb{C})$ of its automophism group and replace the real parameters multiplying those components transforming in spinorial representations with Grassmann variables, i.e define Jordan superfields and study their symmetry superalgebras. The reason for studying Jordan algebras of rank three was due to the fact that the symmetry groups of the exceptional Jordan algebra are all  exceptional groups. Furthermore  the Tits construction relates the other simple Jordan algebras of degree three to the exceptional Lie algebras as well. Therefore, the best hope of constructing the super analogs of exceptional Lie algebras seemed to be via the supersymmetric deformations of Jordan algebras of degree three and  the related symmetry superalgebras.
Implicitly what I was trying to do was to define what was later introduced and classified in the mathematics literature as Jordan superalgebras by Kac \cite{MR0498755} and study their symmetry superalgebras.

\begin{table}
\begin{center}
\begin{tabular}{|c|c|c|c|c|}
 \hline
$J $& $J_{3}^{\mathbb R}$&~$J_{3}^{\mathbb C}$&~$J_{3}^{\mathbb H}$& ~$J_{3}^{\mathbb O}$~\\
\hline
&&&& \\
$Rot(J)$ & $SO(3)$ & $SU(3)$  & $USp(6)$  & $F_4$ \\
&~&~&~&~\\
 $Lor(J)$ & $SL(3,\mathbb{R})$  & $SL(3,\mathbb{C})$& $SU^*(6)$ & $E_{6(-26)}$ \\
&~&~&~&~\\
$Conf(J)$ & $Sp(6,\mathbb{R})$ & $SU(3,3) $ & $ SO^*(12) $ & $E_{7(-25)}$  \\
\hline
\end{tabular}
\caption{\label{euclidean} Simple Euclidean Jordan algebras of degree 3 and their rotation ( automorphism), "Lorentz" ( reduced structure) and "Conformal" ( linear fractional) groups.
The symbols ${\mathbb R}$, ${\mathbb C}$, ${\mathbb H}$, ${\mathbb O}$
represent the four division algebras.
$J_3^{\mathbb A}$ denotes a Jordan algebra of $3\times 3$ hermitian matrices over $\mathbb{A}$.}
\end{center}
\end{table}
 \begin{table}

\begin{center}
\begin{tabular}{|c|c|c|c|c|}
 \hline
$J $& $J_{3}^{\mathbb R}$&~$J_{3}^{{\mathbb C}_s}$&~$J_{3}^{{\mathbb H}_s}$& ~$J_{3}^{{\mathbb O}_s}$~\\
\hline
&&&& \\
$Rot(J)$ & $SO(3)$ & $SL(3,R)$  & $Sp(6,R)$  & $F_{4(4)}$ \\
&~&~&~&~\\
 $Lor(J)$ & $SL(3,\mathbb{R})$  & $SL(3,R)\times SL(3,R)$& $SL(6,R)$ & $E_{6(6)}$ \\
&~&~&~&~\\
$Conf(J)$ & $Sp(6,\mathbb{R})$ & $SL(6,R) $ & $ SO(6,6) $ & $E_{7(7)}$  \\
\hline
\end{tabular}

\end{center}
\caption{\label{split} Simple split  Jordan algebras of degree 3 and their rotation ( automorphism), "Lorentz" ( reduced structure) and "Conformal" ( linear fractional) groups.
The symbols ${\mathbb R}$, ${\mathbb C}_s$, ${\mathbb H}_s$, ${\mathbb O}_s$
represent the split forms of composition algebras.}
\end{table}


\subsection{Unified construction of Lie and Lie super algebras}
There is another construction of the Lie algebras of the Magic Square over the tensor products of two composition algebras whose symmetry is manifest.
and  which we give in Table \ref{kantormagic} in its compact form.
 This is due to Kantor \cite{MR0349779} and is a special case of his general construction of Lie algebras over triple systems which generalizes the Tits-Kantor-Koecher construction of Lie algebras over Jordan triple systems as well as  the construction of exceptional Lie algebras over Freudenthal triple systems.

\begin{table}
	\begin{displaymath}
	\begin{array}{|c|c|c|c|c|}
	\hline
	~ & \mathbb{R} &\mathbb{C} &\mathbb{H}&~~~\mathbb{O}~~~ \\ \hline
	&&&& \\
~	\mathbb{R}~ & SO(3) & SU(3) & ~ USp(6) ~& F_4 \\
	\hline
	&&&& \\
	\mathbb{C} & SU(3) & SU(3)\times SU(3)  & SU(6) & E_6 \\
	\hline
	&&&& \\
	\mathbb{H} & USp(6)& SU(6)  & SO(12) & E_7 \\
	\hline
	&&&&     \\
	\mathbb{O} &F_4& E_6  & E_7 & E_8 \\
	\hline
	\end{array}
	\end{displaymath}
	\caption{ \label{kantormagic} Magic Square constructed over the tensor products of composition algebras due to Kantor. }
\end{table}
Kantor's  construction of the Lie algebras of the Magic Square uses their 5-graded decomposition:

 \eq
 L= L^{-2} \oplus L^{-1} \oplus L^0 \oplus L^1 \oplus L^2
 \en
 such that the elements of the grade $+1$ subspace are labelled by the elements of the underlying triple system which is defined over the tensor product of two composition algebras.

In a joint work with Bars we extended Kantor's general construction to a unified construction of Lie and Lie superalgebras over triple systems\cite{Bars:1978yx}\footnote{We should note that in recent literature on $M2$ branes triple systems are referred to as three algebras. See \cite{Palmkvist:2009qq} for a study of the relations between M2 brane three algebras and triple systems as studied by Kantor, Bars and G\"unaydin.}. In the Bars-G\"unaydin approach the generators in grade $\pm 1$ subspaces of the Lie algebra $L$ in the five grading  go over to odd generators of the corresponding Lie superalgebra.  
For the construction of the Lie algebras over triple  systems defined by tensor products of associative  composition algebras the construction of the corresponding Lie superalgebras is  achieved by taking the parameters multiplying the elements of the triple system to be anticommuting Grassmann parameters. However for the construction of the exceptional Lie algebras $F_4, E_6 , E_7$ and $E_8$  over the tensor product of octonions with other division algebras the corresponding super Jacobi identities fail to be satisfied. By the time of the work of Bars and Gunaydin,  Kac had already classified all simple Lie superalgebras, which included the exceptional Lie superalgebra $F(4)$ whose compact form has the even subgroup $SO(7)\times SU(2)$. The  octonionic construction of $F(4)$  was given later in \cite{Sudbery:1983rb} and its construction using  the approach of Bars and  Gunaydin, but with a modified triple product was given in \cite{MR1961174}.
\begin{table}
\begin{displaymath}
\begin{array}{|c|c|c|c|c|}
\hline
~ & \mathbb{R} & \mathbb{C} &\mathbb{H}&\mathbb{O} \\ \hline
&&&& \\
\mathbb{R} & OSp(1/2) & SU(2/1)& OSp(4/2) & F(4)\\
\hline
&&&& \\
 \mathbb{C} & SU(1/2) & SU(1/2)\times SU(2/1)   & SU(4/2) & ? \\
 \hline
 &&&& \\
  \mathbb{H} & OSp(4/2)& SU(4/2) & OSp(4/8) &  ? \\
 \hline
 &&&&     \\
  \mathbb{O} &F(4)& ?  & ?& ? \\
 \hline
\end{array}
\end{displaymath}
\caption{\label{supermagic} Super Magic Square constructed over super triple systems defined by tensor products of two composition algebras. Note that the super analogs of $E_6,E_7$ and $E_8$\ do not exist as simple Lie superalgebras over the field of real or complex numbers.  }
\end{table}

However there do not exist any Lie superalgebras corresponding to the E-series over the reals $\mathbb{R}$ or complex numbers $\mathbb{C}$. We give  the super Magic Square corresponding to the super counterpart of Kantor's construction of the Magic Square in Table \ref{supermagic}.

\subsection{
Appearance of exceptional groups in maximal supergravity: } 
In 1978 Cremmer and Julia constructed the maximal $N=8$ supergravity in four dimensions and showed that it has the split exceptional group $E_{7(7)}$ as a global symmetry group on shell \cite{Cremmer:1978ds,Cremmer:1979up}. $28$ vector field strengths and their magnetic duals form the 56 dimensional linear representation of $E_{7(7)}$ and the 70 scalar fields of the theory parametrize the symmetric space $E_{7(7)}/SU(8)$ .    In five dimensions the maximal supergravity has $E_{6(6)}$ as a global symmetry of the Lagrangian. The vector fields of the $5d$ theory transform in the 27 dimensional  representation of $E_{6(6)}$ and the scalar manifold is the symmetric space $E_{6(6)}/USp(8)$. In three dimensions all the bosonic physical degrees of freedom of maximal sugra can be dualized to scalar fields which parameterize the symmetric space $E_{8(8)}/SO(16)$ and the global symmetry becomes the largest exceptional group $E_{8(8)}$. We tabulate these facts  in Table \ref{n8sugra}.
\begin{table}
\begin{displaymath}
\begin{array}{|c|c|c|c|}
\hline
d & \# \, vector \,\,fields  & U\,duality \,\,group & scalar \,\, manifold  \\ \hline
&&& \\
d=5 & 27 & E_{6(6)} & \frac{E_{6(6)}}{USp(8)} \\
\hline
&&& \\
d=4 & 28 & E_{7(7)}   & \frac{E_{7(7)}}{SU(8)}  \\
 \hline
 &&& \\
d=3 & - & E_{8(8)}  & \frac{E_{8(8)}}{SO(16)}  \\
 \hline
\end{array}
\end{displaymath}
\caption{ \label{n8sugra} Global symmetries of maximal supergravity in $5,4$ and $3$ dimensions. Note that $E_{6(6)}$ and $E_{7(7)}$ are the Lorentz and conformal groups of the split exceptional Jordan algebra $J_3^{\mathbb{O}_s}$.}
\end{table}
\subsection{ Supersymmetry and K\"ahler geometry}
 Another one of  Bruno Zumino's seminal  papers titled " Supersymmetry and K\"ahler Manifolds" was published in \cite{Zumino:1979et} about a year  after the $N=8$ supergravity theory was constructed. In this paper  he showed that the scalar manifolds of  $N=1$ supersymmetric  non-linear sigma models in $d=4$ must be  K\"ahler manifolds. He was certainly aware of the importance of this result as he remarked  " ..... supersymmetry implies a metric of the K\"ahler type, a fact that we find most remarkable." Indeed this result turned out to be the tip of an iceberg of remarkable connections between geometry and supersymmetry that were uncovered later and which are still continuing today. 

His results were extended to  $N=2$ supersymmetric sigma models by Freedman and Alvarez-Gaume  who showed that their scalar fields must parameterize an Hyper-K\"ahler manifold\cite{Alvarez-Gaume:1981exv}. Later
 Bagger and Witten showed that the scalar manifolds of the theories describing the coupling of $N=2$ hypermultiplets to $N=2$ supergravity must be quaternionic manifolds \cite{Bagger:1983tt}.

\section{ $N=2$ Maxwell-Einstein supergravity theories, Jordan algebras and the Magic Square}

Another example of deep connections between supersymmetry, geometry and non-associative algebras  was uncovered in my work with Sierra and Townsend on Maxwell-Einstein supergravity theories (MESGT)  with eight supersymmetries \cite{Gunaydin:1983rk,Gunaydin:1983bi,Gunaydin:1984ak,Gunaydin:1986fg} .
$N=2$ Maxwell-Einstein supergravity theories describe the coupling of vector multiplets to $N=2$ supergravity.  Bosonic part of the five dimensional  $N=2$ MESGT Lagrangian takes on a very simple form
\begin{eqnarray}
e^{-1}\mathcal{L} &=& -\frac{1}{2} R
-\frac{1}{4}{\stackrel{\circ}{a}}_{IJ} F_{\mu\nu}^{I} F^{J\mu\nu}-
 \frac{1}{2} g_{xy}(\partial_{\mu}\varphi^{x})
(\partial^{\mu} \varphi^{y})+\\ \nn
 && + \frac{e^{-1}}{6\sqrt{6}}  \, C_{IJK} \, \varepsilon^{\mu\nu\rho\sigma\lambda}
 F_{\mu\nu}^{I}F_{\rho\sigma}^{J}A_{\lambda}^{K}
\end{eqnarray}
 It describes the coupling of $(n_V-1)$ vector multiplets  $( A^a_\mu, \lambda^{ai} , \varphi^a )$ where $a,b,..=1,2,...,(n_V-1)$  to $N=2$ supergravity whose multiplet $ (g_{\mu\nu}, \psi_\mu^{i}, A_\mu^0 )$ has one vector field which is the bare graviphoton. Vectors from the graviton multiplet and vector multiplets are combined  $A^I=(A^0,A^a)$ where 
 $ I ,J,K= 0,1,\ldots, (n_V-1) $. The indices $i,j,..=1,2 $
denote the $SU(2)$ R-symmetry indices and the indices $x,y,..$ and $a,b,c,...  =1,\ldots, (n_V-1)  $ denote the curved and flat indices on the scalar manifold with a metric $g_{xy}$, respectively. The indices  
$\mu, \nu,... = 0,1,2,3,4 $ are the spacetime coordinate indices. 
 $e$ and $R$ denote the f\"unfbein determinant and
scalar curvature of spacetime, respectively. $F_{\mu\nu}^{I}$ are field strengths
of the vector fields $A_{\mu}^{I}$. 
 $\stackrel{\circ}{a}_{IJ}$ is the metric of the kinetic energy term of vector fields that depends on the scalar fields $\varphi^{x}$. On the other hand, the
completely symmetric tensor $C_{IJK}$ is constant as required by local
 Abelian gauge symmetries of vector fields.

Perhaps the most remarkable result of my work with Sierra and Towsend on the $5d$ MESGTs was to prove that they are uniquely determined by the constant symmetric tensor  $C_{IJK}$\cite{Gunaydin:1983rk,Gunaydin:1983bi,Gunaydin:1984yj}. The only physical constraint on the C-tensor is that it lead
 to positive definite kinetic energy terms of the fields of the theory and we gave the most general form of this tensor in what came to be called the canonical basis. This result impacted greatly the subsequent work on matter coupled supergravities with eight supercharges \footnote{See  \cite{Lauria:2020rhc} and the extensive list of references therein.} as well as in string theory when it was shown that the components of the C-tensor can be identified with the intersection numbers of Calabi-Yau manifolds when M-theory or type II superstring is compactified to five  or four dimensions resulting in an $N=2$ MESGT coupled to hypermultiplets\cite{Papadopoulos:1995da,Cadavid:1995bk,Ferrara:1996hh}. 
 
Furthermore  we showed that for those $5d$ MESGTs whose  scalar manifolds are symmetric spaces $G/H$ such that $G$ is a symmetry of the Lagrangian the corresponding tensors $C_{IJK}$   are given by the  norm (determinant) $\mathcal{N}_3$ of  Euclidean Jordan algebras $J$ of degree three \footnote{A  Euclidean Jordan algebra $J$\ is characterised by the property that  $ X^2 +Y^2 =0 $ implies that $ X=Y=0 \, \, \forall \,  \, X, \, Y \,  \in \,  J $.}
\[ \mathcal{N}_3(J) = C_{IJK} h^I h^J h^K \]
and hence their classification follows from the classification of Euclidean Jordan algebras $J$ of degree three. Their scalar manifolds in $d=5$ are given by  \[ \mathcal{M}_5 = \frac{Lor(J)}{Rot(J)} \]
where $Lor(J)$ and $Rot(J)$ denote the Lorentz and rotation group of $J$ as explained above. 

There exist unified $N=2$  Maxwell-Einstein Supergravity theories in five dimensions. They are defined as theories with a simple U-duality group $G$ such that all the  vector fields including the graviphoton transform in an irreducible representation of $G$.  By a combination of supersymmetry and simple noncompact global symmetry of the Lagrangian one can map all the fields in these theories into each other.  I recall how Bruno Zumino was surprised when I told him about the existence of such unified matter coupled supergravity theories in five dimensions. 

For MESGTs with symmetric  scalar manifolds there are only four such unified  MESGTs. They are defined by the four simple Euclidean Jordan algebras $J_3^{\mathbb{A}}$ of $3 \times 3$ Hermitian matrices over the four division algebras $\mathbb{R},\mathbb{C},\mathbb{H}$ and $\mathbb{O}$ and  describe the coupling of 5, 8, 14 and 26 vector multiplets to supergravity. 
Scalar manifolds $\mathcal{M}_5 $ of 5d MESGTs defined by simple Euclidean Jordan algebras of degree three are the symmetric spaces 

\begin{eqnarray}
 J \quad &=& \qquad J_{3}^{\mathbb{R}} \qquad \qquad  \qquad J_{3}^{\mathbb{C}} \qquad \qquad \qquad J_{3}^{\mathbb{H}} \qquad \qquad \qquad  J_{3}^{\mathbb{O}} \nonumber \\
 \mathcal{M}_5 &=& SL(3,\mathbb{R})/
SO(3)
 \quad SL(3,\mathbb{C})/
SU(3) \quad SU^{*}(6)/
USp(6) \quad
 E_{6(-26)}/
F_{4} \nonumber
\end{eqnarray}

 On the other hand the generic Jordan family of MESGTs in $d=5$ defined by non-simple Jordan algebras $J= \Gamma_{(1,n-1)} \oplus \mathbb{R} $ has  the scalar manifolds
\eq \left(SO(n-1,1) \times SO(1,1)\right) / SO(n-1) \en
and are not unified MESGTs since their full U-duality groups are not simple.  

The global symmetry groups and the scalar manifolds of the  $N=2$ MESGTs defined by Euclidean Jordan algebras of degree three in four and three dimensions were determined in \cite{Gunaydin:1983rk,Gunaydin:1983bi,Gunaydin:1984ak} and provided the first examples of what came later to be called R-map and C-map. 
Symmetry groups of the theories defined by {\it simple} Euclidean Jordan algebras of degree three  in 5,4 and 3 dimensions correspond precisely to the groups of the Magic Square of Freudenthal, Rozenfeld and Tits \footnote{ See the review article of Freudenthal \cite{MR0170974} for the original references on the Magic Square.} and hence they are referred to as magical supergravity theories. 

 In addition to the four magical supergravity theories there exist three infinite families of unified MESGT's in five dimensions whose scalar manifolds are neither symmetric nor homogeneous!  They are defined by Lorentzian Jordan algebras $J_{(1,n-1)}^{\mathbb{A}} $ of degree $n \geq 3$ generated by $n\times n$ matrices over the division algebras $\mathbb{A}= \mathbb{R},\mathbb{C},\mathbb{H}$ that are Hermitian with respect to the Lorentzian metric $\eta$ \cite{Gunaydin:2003yx}
\[  (X\eta )^\dagger = X \eta \]
where $\eta =diag(-1,+1,+1,...,+1)$.   The C-tensors of these theories are given by structure constants of the traceless elements of $J_{(1,n-1)}^{\mathbb{A}} $. Remarkably the unified MESGTs defined  by Lorentzian Jordan algebras $J^{\mathbb{R}}_{(1,3)}, J^{\mathbb{C}}_{(1,3)}, J^{\mathbb{H}}_{(1,3)}$  coincide with the three magical supergravity theories defined by the Euclidean Jordan algebras $J_3^{\mathbb{A}}$ with $\mathbb{A}= \mathbb{C},\mathbb{H},\mathbb{O}$, respectively. 

 Under dimensional reduction of $5d$ magical MESGTs  to four dimensions the geometry of the scalar manifold becomes special K\"ahler and the resulting bosonic Lagrangian can be written as  
\begin{eqnarray}
  \mathcal{L}^{(4)} = -\frac{1}{2}R  -
   g_{I\bar{J}}    (\partial_{\mu}z^{I})(\partial^{\mu} \bar{z}^{J})
   \nonumber 
 +\frac{1}{4}\textrm{Im}(\mathcal{N}_{AB})F_{\mu\nu}^{A}F^{\mu\nu B}-\frac{1}{8}
   \textrm{Re} (\mathcal{N}_{AB})\epsilon^{\mu\nu\rho\sigma}
   F_{\mu\nu}^{A}F_{\rho\sigma}^{B}
\end{eqnarray}

For theories defined by Euclidean Jordan algebras of degree three the scalar manifolds of four  dimensional $N=2$ MESGTs are of the form \[ \mathcal{M}_4=\frac{Conf(J)}{\widetilde{Lor}(J) \times U(1) } \] 
where $Conf(J)$ denotes the conformal group of the underlying Jordan algebra $J$ and $\widetilde{Lor}(J) $ denotes  the compact real form of the Lorentz group of $J$.

One-to-one correspondence between vector fields  $A^{\mu}_I  $ and the elements of underlying Jordan algebra $J$ of degree three in $5d$ MESGT goes over to a one-to-one correspondence between the elements of the field strengths and their magnetic duals $F_{\mu\nu}^A \oplus \Tilde{F}_{\mu\nu}^A $  and the elements of the Freudenthal triple system (FTS) $\mathcal{F}(J)$ defined by $J$ which can be represented formally as a $2\times 2$ matrix  \cite{Ferrara:1997uz,Gunaydin:2000xr,Gunaydin:2005zz} :
\begin{equation} \normalsize \mathcal{F}(J) \ni  X=
\begin{array}{|ccc|c|ccc|}
\mathbb{R}  &  & \mathbf{J} & &  F^0_{\mu\nu} &  &  F^I_{\mu\nu}\\
&  &  & \Leftrightarrow & &  & \\
\mathbf{\Tilde{J}}  &  & \mathbb{R}  && \Tilde{F}^I_{\mu\nu} &  & \Tilde{F}^0_{\mu\nu}
\end{array}
\end{equation}
where $\alpha, \beta \in \mathbb{R}$ and $\mathbf{J}$ and $\tilde{\mathbf{J}}$ are elements of $J$.  $F^0_{\mu\nu}$ and $\Tilde{F}^0_{\mu\nu}$ denotes the field strength of the vector field coming from the $5d$ graviton and its dual. 

 Scalar manifolds  of 4d magical MESGTs are the following Hermitian symmetric spaces   :
\begin{eqnarray} 
 J \quad &= &\qquad J_{3}^{\mathbb{R}} \qquad \qquad  J_{3}^{\mathbb{C}} \qquad \qquad \qquad J_{3}^{\mathbb{H}} \qquad \qquad   J_{3}^{\mathbb{O}} \nonumber \\
 \mathcal{M}_4 &=& \frac{Sp(6,\mathbb{R})}{
U(3)}
 \quad \frac{SU(3,3)}{
S(U(3)\times U(3))} \quad \frac{SO^{*}(12)}{
U(6)} \quad
 \frac{E_{7(-25)}}{
E_{6} \times U(1)} \nonumber
\end{eqnarray}

 $N=2$ MESGTs reduce to $N=4$ supersymmetric sigma models coupled to supergravity in $d=3$. The target manifolds of magical supergravity theories  in $d=3$ are the exceptional  quaternionic symmetric spaces:
 
 \[  \frac{F_{4(4)}}{USp(6)\times USp(2)} \,,\, \frac{E_{6(2)}}{SU(6)\times SU(2)}
  \, , \, \frac{E_{7(5)}}{SO(12)\times SU(2)} \, ,\, \frac{E_{8(-24)}}{E_7 \times SU(2)} \] 
  
   The generic Jordan family of MESGTs reduced to $d=3$ have the target spaces:
   \[ \frac{SO(n+2,4)}{SO(n+2)\times SO(4)} \] and the pure $N=2$ , $5d$ supergravity reduces to $N=4$ sigma model with target space $\frac{G_{2(2)}}{SO(4)}$.
 
\section{ Conformal and quasiconformal  groups of Jordan algebras as spectrum generating symmetry groups $N=2$ Maxwell-Einstein supergravity theories}

U-duality orbits of extremal , spherically symmetric stationary black hole solutions of $N=2$ MESGTs with symmetric target spaces as well as maximal supergravity in $d=5$ and $d=4$  were first studied in \cite{Ferrara:1997uz}.  This work led to the proposal that $4d$ U-duality groups must act as spectrum generating conformal groups of the underlying Jordan algebras that define the corresponding $5d$ supergravity theories\cite{Ferrara:1997uz,Gunaydin:2000xr,Gunaydin:2004ku,Gunaydin:2005gd}. $Conf[J]$ leaves invariant  light-like separations  with respect to a cubic distance function $\mathcal{N}_3(J_1 -J_2)$ and admits a 3-grading with respect to their Lorentz subgroups
\[ Conf[J] = K_J \oplus Lor(J) \times \mathcal{D} \oplus T_J \]
 $Lor(J)$ is the $5D$  U-duality group that leaves the cubic norm invariant.

 The degeneracies of BPS black holes of  exceptional supergravity in five dimensions was studied more recently in my work with Kidambi\cite{Gunaydin:2022qvw}. Since charges of black holes take on integer values and can be represented by elements of the exceptional Jordan algebra over integral octonions, labelled as $J_3^{\mathbb{O}}(\mathcal{R})$, we defined the quantum degeneracy as the number of distinct elements of $J_3^{\mathbb{O}}(\mathcal{R})$ with a given set of invariant labels of the discrete U-duality group  $E_{6(-26)}(\mathbb{Z})$. Using the remarkable results of Elkies and Gross \cite{MR1411589,MR1845183} we showed that the quantum degeneracies of charge states of rank one  BPS black holes are given by the Fourier coefficients of unique singular modular form  $E_4(Z)$ of weight 4  of the discrete arithmetic  subgroup  $E_{7(-25)}(\mathbb{Z})$ of the conformal group $E_{7(-25)}$ of the exceptional Jordan algebra\cite{MR1216126}. Similarly using the results of \cite{MR1437509,MR1195510,MR2478255} we showed that quantum degeneracies of rank two BPS black holes are similarly given by the singular modular form $E_8(Z)=(E_4(Z))^2$ of weight 8 of $E_{7(-25)}(\mathbb{Z})$ which is also unique.  These results support the proposal that the conformal groups of Jordan algebras of degree three act as spectrum generating groups of the $5d$ MESGTs defined by them. For the exceptional supergravity the conformal group is $E_{7(-25)}$ which is isomorphic to the U-duality group of $4d$ exceptional supergravity theory. For our work we used some of the results of Bruno Zumino and collaborators on the orbits of black holes including rank one and rank two small BPS black holes (zero area) in one of his last papers \cite{Cerchiai:2010xv} extending earlier work of \cite{Ferrara:1997uz,Ferrara:2006xx}.

 The above proposal naturally suggested the question whether the $3D$ U-duality groups can act as spectrum generating  symmetry groups  of corresponding  $4D$ supergravity theories which was investigated by Gunaydin, Koepsell and Nicolai (GKN)  \cite{Gunaydin:2000xr} and led to the discovery of novel geometric quasiconformal realizations of  noncompact groups , in particular those of $E_{8(8)}$ and $E_{8(-24)}$ which are the U-duality groups of maximal and exceptional supergravity theories in 3 dimensions. 

   Starting from the fact that no conformal realization for any real forms of $E_8$, $G_2$ and $F_4$ exists since their Lie algebras do not admit  3-grading  with respect to a subalgebra of maximal rank GKN considered 5-graded decompositions of  simple  Lie algebras with respect to a subalgebra of maximal rank \[ \mathfrak{g} = \mathfrak{g}^{-2} \oplus \mathfrak{g}^{-1} \oplus \mathfrak{g}^0 \oplus \mathfrak{g}^{+1} \oplus \mathfrak{g}^{+2} \] such that the grade $\pm 2$ subspaces are one-dimensional\footnote{ All simple Lie algebras admit such a 5-graded decomposition except for $sl(2)$ .}
   \eq
  \mathfrak{g}  =  \tilde{K}  \oplus \tilde{U}_A  \oplus [S_{(AB)}+ \Delta ]  \oplus U_A  \oplus K \en
 $ A,B,C...= 1,...2N $ and $( K, \Delta , \tilde{K} )$ form an $sl(2)$ subalgebra.
 The Lie algebras with such a 5-grading can be constructed over Freudenthal triple systems (FTS) such that the indices $A$ in $U_A \in  \mathfrak{g}^{+1} $ are labelled by  elements of a FTS  $ \mathcal{F}$ and all the commutation relations can be expressed in terms of Freudenthal triple product $(A,B,C)$ and the skew-symmetric invariant form $\langle A | B \rangle$ of the FTS. 

Non-linear quasiconformal realization of   $E_{8(8)}$ as formulated by GKN uses its 5-grading with respect to $E_{7(7)}$ 
\begin{eqnarray}
E_{8(8)}  & = & \mathfrak{1}_{-2}  \oplus \mathfrak{56}_{-1}  \oplus
       E_{7(7)} + SO(1,1)  \oplus \mathfrak{56}_{+1}
    \oplus \mathfrak{1}_{+2}  \nn \\
  \mathfrak{g} & = & \tilde{K}  \oplus \tilde{U}_A  \oplus [S_{(AB)}+ \Delta ]  \oplus  U_A  \oplus K \nn
\end{eqnarray}
One defines its action over a  space $\mathcal{T}$ coordinatized by the
elements $X$ of  the exceptional FTS $\mathcal{F}(J_3^{\mathbb{O}_S})$ plus  an extra singlet variable $x$ corresponding to grade $+1$ and grade $+2$ subspaces
\[  \mathfrak{56}_{+1}
     \oplus \mathfrak{1}_{+2} \, \Leftrightarrow (X,x) \in  \mathcal{T} \] 
as follows 
\begin{equation*}
\begin{split}
  \begin{aligned}
      K\left(X\right) &= 0 , \\
      K\left(x\right) &= 2,\,
  \end{aligned}
  & \quad
  \begin{aligned}
     U_A \left(X\right) &= A, \\
     U_A\left(x\right) &= \left< A, X\right>,
  \end{aligned}
   \quad
   \begin{aligned}
      S_{AB}\left(X\right) &= \left( A, B, X\right), \\
      S_{AB}\left(x\right) &= 2 \left< A, B\right> x,
   \end{aligned}
 \\
 &\begin{aligned}
    \Tilde{U}_A\left(X\right) &= \frac{1}{2} \left(X, A, X\right) - A x \\
    \Tilde{U}_A\left(x\right) &= -\frac{1}{6} \left< \left(X, X, X\right), A \right> + \left< X, A\right> x
 \end{aligned}
 \\
 &\begin{aligned}
    \Tilde{K}\left(X\right) &= -\frac{1}{6} \,  \left(X,X,X\right) +  X x \\
    \Tilde{K}\left(x\right) &= \frac{1}{6} \,  \left< \left(X, X, X\right), X \right> + 2\,  \, x^2
 \end{aligned}
\end{split}
\end{equation*}
 where $ \langle (X,X,X),X\rangle $ is  the quartic invariant of $E_{7(7)}$ and 
 $A,B$ are some constant elements of $ \mathcal{F}(J_3^{\mathbb{O}_S}) $.


To make manifest the geometric meaning of the quasiconformal action of the Lie algebra $E_{8(8)}$  on the 57 dimensional space $\mathcal{T}$ GKN defined  a  quartic norm $ \cN_4(\cX) $ of vectors  $\cX =(X,x)\in \mathcal{T}$ as
\eq
 \cN_4(\cX) := \mathcal{Q}_4(X) - x^2
\en
 where $\mathcal{Q}_4(X)$ is the quartic norm  of the underlying Freudenthal system , which in this case corresponds to the quartic invariant of $E_7$. 
Using this quartic norm one then defines   a  quartic ``distance'' function between any two points $\cX=(X,x)$ and
$\cY=(Y,y) $ in $\mathcal{T}$ as  
\[
d(\cX,\cY):= \color{black}  \cN_4 ( \gd(\cX,\cY)
\] 
  $\gd(\cX,\cY)$ is the ``symplectic'' difference of $\cX $ and $\cY$ :
\[
  \gd(\cX,\cY) := (X-Y,x-y+ \color{black} \langle X, Y \rangle  )= - \gd(\cY , \cX)
\]
Remarkably one finds that light-like separations $
  d(\cX,\cY)=0
$
 are left invariant under  the nonlinear quasiconformal group action. In other words $E_{8(8)}$ is the invariance group  of a "quartic light-cone" in 57 dimensions. Quasiconformal realizations of various real forms of $E_8$ represent the only known geometric realizations of $E_8$. 
Quasiconformal realizations exist for all simple Lie algebras  as well as  Lie superalgebras \cite{Gunaydin:2000xr,Gunaydin:2001bt,Gunaydin:2006vz,Fernando:2010dp,Fernando:2009fq,Fernando:2014pya}.

The quantization of the geometric quasiconformal realizations of non-compact Lie algebras and Lie superalgebras leads directly to their minimal unitary representations and provides a unified aproach to the theory of minimal unitary realizations of Lie algebras and Lie superalgebras. 
Furthermore quasiconformal realizations also provides a geometric framework for the explicit construction of quaternionic series representations of noncompact groups that were introduced aand studied algebraically by Gross and Wallach \cite{gross_wallach}. For the groups $SU(2,1)$ and $G_{2(2)}$ this construction was carried out in great detail in \cite{Gunaydin:2007qq}. The K-type spectrum of the quaternionic series representations is generated by the action of the noncompact generators on the spherical vectors of their maximal compact subgroups.   The spherical vectors of the quaternionic noncompact groups $F_{4(4)},E_{6(2)}, E_{7(-5)}, E_{8(-24)}$ and $SO(n+4,4)$ that are the U-duality groups of $3d$ MESGTs defined by Jordan algebras of degree 3 were determined explicitly in terms of their C-tensors in \cite{Gunaydin:2009dq}.

The proposal that three
dimensional U-duality groups must act as spectrum generating quasiconformal
groups of  four dimensional $N=2$ MESGTs was realized concretely for spherically
symmetric stationary BPS black holes   in 
\cite{Gunaydin:2005mx,Gunaydin:2007bg,Gunaydin:2007qq}.   The framework presented in these papers is  based on the fact that the
attractor equations for a spherically symmetric stationary black hole of
four dimensional supergravity theory are equivalent to the equations for  geodesic
motion of a fiducial particle on the moduli space of
$3d$ supergravity obtained by reduction on a
time-like circle \cite{Breitenlohner:1987dg}. These moduli spaces are para-quaternionic spaces of the form
\begin{equation}
\mathcal{M}_3^*(J) = \frac{QConf(J)}{ Conf(J) \times SU(1,1)}
\end{equation}
for $N=2$ MESGTs theories defined by Euclidean Jordan algebras $J$ of degree 3. 

The quantum mechanical wave-functions obtained by  quantization of the motion of fiducial particle on $\mathcal{M}_3^*$ form  the basis of a unitary
representation of the isometry group $G_3$ of $\mathcal{M}_3^*$. For $N=2$ MESGTs  defined by Euclidean Jordan algebras $J$  of degree three . the isometry group is simply the quasiconformal group $QConf(J)$ associated with $J$.

 \section{Exceptional $N=2$ supergravity  versus the maximal   $N=8$ supergravity and the exceptional Jordan algebra}

 The exceptional $N=2$ supergravity in five dimensions  is defined by the C-tensor $C_{IJK}$ given by the norm form of the  exceptional Jordan algebra $J_3^{\mathbb{O}}$ of $3 \times 3$ Hermitian matrices over real octonions $\mathbb{O}$. The C-tensor describes the $F ^I \wedge F^J  \wedge A^K $ coupling of vector fields in five dimensions. Its invariance  group  $E_{6(-26)}$ with maximal compact subgroup $F_4$ is the global symmetry group of the Lagrangian of the $5d$ exceptional supergravity.  On the other hand the 
 C-tensor $C_{IJK}$ of  maximal $5d$ , $N=8$  supergravity that describes the $F ^I \wedge F^J  \wedge A^K $ coupling of vector fields can be identified with the symmetric tensor given by the cubic norm of the {\it split}  exceptional Jordan algebra
$J_3^{\mathbb{O}_s}$  defined over split octonions $\mathbb{O}_s$. Its  invariance group  is $E_{6(6)}$ with maximal compact subgroup $USp(8)$. The work of GST showed how to construct the exceptional $N=2$ supergravity in five dimensions starting from the C-tensor given by the norm of exceptional Jordan algebra over real octonions. Whether one can construct  the five dimensional $N=8$ supergravity starting from the C-tensor given by the cubic norm of the exceptional Jordan algebra over split octonions is an interesting open problem. It is remarkable that both the maximal supergravity and the exceptional supergravity theories are related to the exceptional Jordan algebra which is the most famous intrinsically non-associative algebra in mathematics.

 In $D=4$ and $D=3$ the exceptional supergravity has $E_{7(-25)}$ and $E_{8(-24)}$ as its U-duality group while the maximal $N=8$ supergravity has $E_{7(7)}$ and $E_{8(8)}$, respectively. 
 The exceptional supergravity and the maximal supergravity has a common subsector which is the $N=2$ magical supergravity described by the quaternionic Jordan algebra whose U-duality symmetry groups are $SU^*(6) , SO^*(12)$ and $E_{7(-5)}$ in five, four and three dimensions , respectively. 
 
Since neither the maximal supergravity nor the exceptional supergravity can be gauged so as to accommodate the standard model gauge group I speculated that the theory we are looking for unification of the standard model with gravity must be one that can be truncated to maximal supergravity as well as to the exceptional supergravity\cite{Gunaydin:1984be}.  
This was before the so-called first string revolution. At the time we knew that maximal supergravity arises as low energy effective theory of type II superstring compactified on six tori. After the first string revolution whether  the octonionic magical supergravity can also arise as low energy effective theory of  superstring theory compactified  on some Calabi-Yau manifold was posed as an open problem in \cite{Gunaydin:1986fg}. 
 Later it was pointed out in \cite{Gunaydin:2009pk} that the four dimensional  quaternionic magical supergravity  corresponds to one of the constructions of a dual pair of theories in Type IIB superstring theory given in \cite{Sen:1995ff}. 

The magical  supergravity theories including the exceptional supergravity coupled to hypermultiplets  in $d=6$ was constructed in \cite{Gunaydin:2010fi}.  Six dimensional exceptional supergravity has the global symmetry $SO(9,1)$ and the maximal supergravity has $SO(5,5)$ as their U-duality groups in six dimensions. 
The  exceptional $N=2$ supergravity in $d=6$ coupled to 28 hypermultiplets parametrizing the coset space $\frac{E_{8(-24)}}{E_7\times SU(2)} $  is free of gravitational anomalies. When reduced to four dimensions it has the scalar manifold
\eq 
\frac{E_{7(-25)}}{E_6 \times U(1)} \times \frac{E_{8(-24)}}{E_7\times SU(2)}
\en
which has the moduli space of FHSV model\cite{Ferrara:1995yx} as a sub-manifold since
\[ SO(10,2) \times SU(1,1)\times SO(12,4)  \subset E_{7(-25)} \times E_{8(-24)} \]
In three dimensions this exceptional theory with exceptional matter has the moduli space \cite{Gunaydin:2009pk}
\eq  \frac{E_{8(-24)}}{E_7\times SU(2)} \times \frac{E_{8(-24)}}{E_7\times SU(2)} \en
.
 
To this date no rigid Calabi-Yau  manifold was found that would lead to the exceptional supergravity  from  M/superstring theory via compactification. 
The focus  shifted to look for a self-mirror Calabi-Yau manifold with $h_{11}=h_{12}=27$  after it was realized that there exists an anomaly free supergravity theory in six dimensions that reduce to the octonionic magical supergravity theory coupled to 28 hypermultiplets with the target space $E_{8(-24)}/E_7\times SU(2) $ in five and four dimensions.
Remarkably there is a very sharp peak at the number of self-mirror Calabi-Yau manifolds with these Hodge numbers , namely there exist 910,113 such toric threefolds. However to show whether the intersection numbers of any of these threefolds coincide with the C-tensor of the exceptional supergravity is currently beyond the available computational tools. For the review of recent status of these searches and the references on the subject  see \cite{Gunaydin:2022qvw}.

The maximal ungauged $N=8$ supergravity is unique and can not be embedded in a higher supergravity theory with a single graviton. Less known is the fact that  the exceptional supergravity can not be embedded in a larger $N=2$ MESGT without breaking its global symmetry group.  This follows from the fact that exceptional Jordan algebra that defines the exceptional supergravity can not be embedded in a larger Jordan algebra. This parallels the fact that the octonionic quantum mechanics defined over the exceptional Jordan algebra $J_3^{\mathbb{O}}$ can not be embedded in a higher quantum mechanics due to the non-Desarguesian nature of the underlying projective  geometry. 

Since the standard model gauge group can not be embedded in any of the gaugings of the maximal supergravity  Ellis, Gaillard
, Maiani and Zumino proposed a  composite scenario with the goal of relating  the maximal $N=8$ supergravity to observation\cite{Ellis:1980tf}. In this scenario one assumes that the local composite $SU(8)$ symmetry of maximal supergravity in four dimensions becomes dynamical at the quantum level and the bound states form a linear unitary representation of the U-duality group $E_{7(7)}$ which is infinite dimensional.  They then speculated that this family unifying dynamical symmetry might be broken down to its $SU(5)\times SU(3)$ subgroup such that one is left with an effective $SU(5)$ GUT with three families of quarks and leptons at low energies. 
This proposal was one of the original motivation of my work with Saclioglu \cite{Gunaydin:1981dc,Gunaydin:1983cc} on the unitary representations of U-duality groups of extended ($N\geq 4$) supergravity theories which led eventually to the general oscillator construction of the unitary representations of non-compact groups \cite{Gunaydin:1981yq}. 

With the discovery of the exceptional supergravity  G\"unaydin, Sierra and Townsend  proposed a similar composite scenario for the exceptional supergravity in which the composite local symmetry $E_6 \times U(1)$ becomes dynamical at the quantum level  \cite{Gunaydin:1983rk,Gunaydin:1984be} leading to an effective $E_6$ GUT.  Unlike $E_{7(7)}$ symmetry  of maximal supergravity,  the U-duality group $E_{7(-25)}$ of the exceptional supergravity admits unitary lowest weight representations. These representations  are intrinsically chiral \cite{Gunaydin:1984be} in the sense that complex representations of $E_6$ subgroup do not appear in complex conjugate pairs in the Hilbert space of an irreducible unitary lowest ( or highest) weight representation of  $E_{7(-25)}$ and hence  their spectra can accommodate   chiral families of quarks and leptons in a composite scenario.  In a recent work I showed that the minimal unitary representation of $E_{7(-25)}$ can accommodate an infinite family of quarks and leptons transforming in the spinor representation $16$  of $SO(10)$ when $E_6$ is broken down to its $SO(10)\times U(1)$  subgroup\cite{Gunaydin:2024ihp}. To be able to relate it to the standard model one has to break $SO(10)$ further to the standard model gauge group $SU(3)\times SU(2) \times U(1)$ such that one is left with 3 families of quarks and leptons at low energies\cite{Gunaydin:2024ihp}.

  \section{Superspaces coordinatized by Jordan superalgebras and the exceptional superspace }

\subsection{
The space-times defined by simple Jordan algebras of degree 3 }

The generalized spacetimes defined by simple Euclidean Jordan algebras $J_3^\mathbb{A}$ of degree three ($\mathbb{A}=\mathbb{R}, \mathbb{C},\mathbb{H},\mathbb{O}$) correspond to extensions of Minkowski space-times with coordinates $ x_m$ 
in the critical dimensions $d=3,4,6,10$ by a dilatonic ($\rho$)
\emph{and} commuting spinorial coordinates ($\xi^\alpha$).
\begin{equation}
\begin{split} \nonumber
   J_3^\mathbb{R}  &\Longleftrightarrow \left(\rho, x_m, \xi^{\alpha} \right) \quad m=0,1,2 \quad \alpha =1,2 \\
   J_3^\mathbb{C}  &\Longleftrightarrow \left(\rho, x_m, \xi^{\alpha} \right) \quad m=0,1,2,3 \quad \alpha =1,2,3,4 \\
   J_3^\mathbb{H}  &\Longleftrightarrow \left(\rho, x_m, \xi^{\alpha} \right) \quad m=0,\ldots,5 \quad \alpha =1,\ldots,8 \\
   J_3^\mathbb{C}  &\Longleftrightarrow \left(\rho, x_m, \xi^{\alpha} \right) \quad m=0,\ldots,9 \quad \alpha =1,\ldots,16
\end{split}
\end{equation}
We can  represent the elements of $J_3^\mathbb{A}$  as 
\eq
\left(\begin{array}{cc}\sigma_m x^m & \xi^\alpha \\ \xi_\alpha  & \sqrt{2} \, \rho \end{array}\right)
\en
where $\sigma_m$ are the Pauli matrices in the critical dimensions and the commuting spinors $\xi^\alpha$ 
are represented by a $2 \times 1$ matrices  over $\mathbb{A}=\mathbb{R}, \mathbb{C},\mathbb{H},\mathbb{O}$.  They correspond to Majorana, Weyl , symplectic Majorana-Weyl and Majorana-Weyl spinors in $d=3,4,6$ and $d=10$ dimensions respectively. The
cubic norm of a ``vector''  with coordinates $X^I=\left(\rho, x_m,
\xi^{\alpha}\right)$ in the generalized space-times is given by
\begin{equation} \nonumber
  \mathcal{V}\left(\rho, x_m, \xi^{\alpha}\right) = C_{IJK}X^IX^JX^K=\sqrt{2} \rho x_m x_n \eta^{mn} + x^m \Bar{\xi} \gamma_m \xi
\end{equation}
The C-tensors of Jordan algebras of degree three satisfy the so-called adjoint identity 
\eq
C^{IJK} \, C_{J(MN} \, C_{PQ)K} = \delta^I_{(M} C_{NPQ)}
\en 
Using the above decomposition of elements of $J_3^\mathbb{A}$ Sierra showed that the adjoint identity implies the Fierz identities for the existence of super Yang-Mills theories in the critical dimensions \cite{Sierra:1986dx}\footnote{ Remarkably, the list of cubic forms that
satisfy the adjoint identity that first appeared in \cite{Gunaydin:1983bi} in the physics literature coincides also with the list of Legendre invariant cubic forms that
were classified by mathematicians two decades later \cite{MR1890194}}. 

The Lorentz groups of the space-times over $J_3^\mathbb{A}$ are
\begin{equation}
   \mathrm{SL}\left(3, \mathbb{R}\right), \quad \nonumber
    \mathrm{SL}\left(3, \mathbb{C}\right), \quad
    \mathrm{SU}^\ast\left(6\right), \quad \text{and}
    \quad \mathrm{E}_{6(-26)}
\end{equation}
respectively, corresponding to the invariance groups of their cubic
norm.
Their Lorentz groups have the Lorentz groups Minkowskian spacetimes in $d=3,4,6,10$ as subgroups:
\begin{equation} \normalsize
\begin{split} \nonumber
    J = \mathbb{R} \oplus J_2^\mathbb{R}  &: \mathrm{SO}(1,1) \times \mathrm{SO}\left(2,1\right) \subset \mathrm{SL}\left(3, \mathbb{R}\right) \cr
    J = \mathbb{R} \oplus J_2^\mathbb{C}  &: \mathrm{SO}(1,1) \times \mathrm{SO}\left(3,1\right) \subset \mathrm{SL}\left(3, \mathbb{C}\right) \cr
     J = \mathbb{R} \oplus J_2^\mathbb{H}  &: \mathrm{SO}(1,1) \times \mathrm{SO}\left(5,1\right) \subset \mathrm{SU}^\ast\left(6\right) \cr
     J = \mathbb{R} \oplus J_2^\mathbb{O}  &: \mathrm{SO}(1,1) \times \mathrm{SO}\left(9,1\right) \subset \mathrm{E}_{6(-26)}
\end{split}
\end{equation}
The conformal groups of generalized spacetimes defined by Euclidean Jordan algebras have  positive energy unitary representations and as a consequence  they admit global invariant causal structures\cite{Mack:2004pv}. However this is not true for spacetimes defined by non-euclidean Jordan algebras, in general. For example the spacetime defined by split exceptional Jordan algebra $J_3^{\mathbb{O}_s} $has the conformal group $E_{7(7)}$ which does not admit positive energy unitary representations. Its Lorentz group $E_{6(6)}$ has the subgroup $SO(5,5)$. The adjoint identities of $J_3^{\mathbb{O}_s} $ imply the analogs of Fierz identities in a ten dimensional space-time with signature $(+++++-----)$.


\subsection{Superspaces defined by Jordan superalgebras}

 Jordan superalgebras $J$ are $Z_{2}$ graded algebras with a  super-commutative product \cite{MR0498755}:

\eq
A \circ B = (-1)^{\alpha \beta} B\circ A, \quad (\alpha=\text{deg}\,A,\,\beta=\text{deg}\,B)
\en
that satisfy the identity 
\bea   ((A\circ B)\circ C)\circ D + (-1)^{\beta \gamma + \beta \gamma + \gamma \delta } 
((A\circ D)\circ C)\circ B + (-1)^{\alpha \beta +\alpha \gamma+\alpha \delta +\gamma \delta}
((B\circ D)\circ C)\circ A  \nonumber \\ =
(A\circ B)\circ (C\circ D)  + (-1)^{\beta \gamma} 
(A\circ C)\circ (C\circ D) + (-1)^{\delta ( \beta + \gamma )} (A\circ D)\circ (B\circ C) \nonumber
\eea
 where $\gamma =\text{deg}\, C$ and $\delta=\text{deg}\, D$ and $\alpha , \beta , \gamma , \delta = 0,1$ . 
 Their even subalgebras generated by degree zero elements are ordinary Jordan algebras. 
One can extend the concept of generalized space-times coordinatized by Jordan algebras to define generalized superspaces coordinatized by Jordan superalgebras such that  their rotation, Lorentz and  conformal superalgebras are identified with  their automorphism, reduced structure
and conformal ( linear fractional)  superalgebras \cite{Gunaydin:1979df,Gunaydin:1989ur,Gunaydin:1992zh}. We give the symmetry superalgebras of superspaces defined by simple Jordan superalgebras in Table \ref{superspaces}. 

 \small
 \begin{table}
\begin{displaymath}
\begin{array}{cccc}
~&~&~&~\\
JX & SRot & SLor & SConf \\
\hline
~&~&~&~\\
JA(m^{2}+n^{2}/2mn) & SU(m/n) &
SU(m/n)\times SU(m/n) & SU(2m/2n) \\
~&~&~&~\\
JBC(r/s) & OSp(m/2n) &
SU(m/2n) & OSp(4n/2m) \\
~&~&~&~\\
JD(m/2n) & OSp(m-1/2n) &
OSp(m/2n) & OSp(m+2/2n) \\
~&~&~&~\\
JP(n^{2}/n^{2}) & P(n-1) &
SU(n/n) & P(2n-1) \\
~&~&~&~\\
JQ(n^{2}/n^{2}) & Q(n-1)\times U(1)_{F} &
Q(n-1)\times Q(n-1) \times U(1)_{F} & Q(2n-1) \\
~&~&~&~\\
JD(2/2)_{\alpha} & OSp(1/2) &
SU(1/2) & D(2,1;\alpha) \\
~&~&~&~\\
\bf JF(6/4) & \bf OSp(1/2)\times OSp(1/2) &
\bf OSp(2/4) & \bf F(4) \\
~&~&~&~\\
JK(1/2) & OSp(1/2) &
SU(1/2) & SU(2/2) \\
~&~&~&~
\end{array}
\end{displaymath}
\caption{\label{superspaces}  Above we list the simple Jordan superalgebras $JX(./.)$ and the super-rotation , super-Lorentz and super-conformal  algebras $SRot , SLor , SConf $ of the superspaces defined by them. The first and second arguments  of $JX(./.)$ indicate the dimensions of bosonic and fermionic subspaces of the superspace defined by it.   Dimensions of bosonic ($r$) and fermionic $(s)$ subspaces of the superspace defined by the Jordan superalgebra $JBC(r/s)$ are $r= \frac{1}{2}m(m+1) +n(2n-1)$ and $s = 2mn$. The exceptional superspace defined by the Jordan superalgebra $JF(6/4)$ and its symmetry super-algebras are indicated in boldface. }
\end{table}

\subsection{ The  Exceptional  Superspace }
 In the list of simple Jordan superalgebras one is truly unique,
namely the exceptional Jordan superalgebra $JF(6/4)$\cite{MR0498755}. It is the
only simple Jordan superalgebra which has no realization in terms
of $Z_{2}$ graded associative supermatrices.  The even ("bosonic") elements of $JF(6/4)$ belonging to grade zero subspace
are denoted as $S,B_{0}$ and $B_{i},~(i=1,2,3,4)$. The odd
elements belonging to the grade one subspace are denoted as
$Q_{\alpha},~ (\alpha=1,2,3,4)$. Their super-commutative Jordan
products are\cite{Gunaydin:1989ur,Gunaydin:1989mgm}:

\begin{eqnarray}
B_{i} \cdot B_{j} & = & -\delta_{ij} B_{0}
\quad  ,   \quad
B_{0} \cdot B_{i}    ~=~ B_{i}\quad , \quad 
B_{0} \cdot  B_{0}   ~=~ B_{0} \\
B_{0} \cdot S   &=& 0=~B_{i} \cdot S \quad , \quad
S \cdot S   ~=~ S   \\
Q_{\alpha} \cdot Q_{\beta}   &=& (i \gamma_{5} \gamma_{i}
C)_{\alpha \beta} B^{i}~+~(\gamma_{5}C)_{\alpha \beta}
(B_{0}-3S)\\
B_{i} \cdot Q_{\alpha}  &=& \frac{i}{2}(\gamma_{i})_{
\alpha \beta} Q_{\beta}
\quad , \quad
B_{0} \cdot Q_{\alpha} = \frac{1}{2}Q_{\alpha}\\
S \cdot Q_{\alpha} &=& \frac{1}{2}Q_{\alpha}
\\
 i,j,\ldots &=& 1,2,3,4 \quad ~;~\alpha,\beta,\ldots~=~1,2,3,4 \nonumber
\end{eqnarray}
 $B_{0}$ and $S$ are the two idempotents and $I = B_{0}+S$ is
the identity element of $JF(6/4)$. The matrices $\gamma_{i}$
are the four-dimensional (Euclidean)  Dirac gamma  matrices and
$C$ is the charge conjugation matrix:
\begin{equation}
\{\gamma_{i},\gamma_{j}\}  = 2 \delta_{ij} \quad , \quad
\gamma_{i} C ~=~ -C \gamma_{i}^{T}
\end{equation}

 Super rotation, super Lorentz and  conformal  Lie superalgebras of  the exceptional superspace coordinatized by $JF(6/4)$ and their even subalgebras  are:
  \eqn
 SRot(JF(6/4))&=&  OSp(1/2) \times OSp(1/2) \supset SU(2)\times SU(2) \\     SLor(JF(6/4))& = &
  OSp(2/4) \supset SO(2) \times USp(2,2)\supset SO(2)\times SO(4) \\
SConf(JF(6/4))&=&
 F(4) \supset SO(5,2) \times SU(2)\supset SO(5)\times SO(2) \times SU(2) \enn
 Hence $F(4)$ is simply the superconformal algebra in five dimensions with an R-symmetry group $SU(2)$ and 16 supercharges. 
 Non-linear action of $F(4)$ on the ten dimensional exceptional superspace can be readily obtained using the quadratic Jordan formulation \cite{Gunaydin:1989ur,Gunaydin:1992zh}.

According to Nahm's classification $d=6$ is the maximal dimension for the existence of superconformal field theories\cite{Nahm:1977tg}!
The exceptional $N=2$ superconformal algebra $F(4)$ in five dimensions can not be embedded in any six dimensional  super conformal algebra $OSp(8^*|2N) \supset SO(6,2) \times USp(2N) $. This can be understood to be a consequence of the exceptionality of the Jordan superalgebra $JF(6/4)$ which can not be embedded in any higher Jordan superalgebra.

Minimal unitary realization of $F(4)$ was obtained via quasiconformal techniques in  \cite{Fernando:2014pya}. The enveloping algebra of the minimal unitary representation of $F(4)$ is the unique higher spin super algebra in five dimensions \cite{ Fernando:2014pya} and the corresponding supersymmetric higher spin theory  was formulated and studied in \cite{Gunaydin:2016amv}.

Early attempts to find the super analogs of exceptional groups for the E-series \cite{Gunaydin:1975mp,Bars:1978yx} yielded negative results  over the fields of real and complex numbers. As  can be seen from Table \ref{supermagic} the only exceptional group in the Magic Square  that has a super analog over the reals or complex numbers if $F_4$. In  preparing for my talk for the Bruno Zumino memorial conference I looked at the mathematics literature to see if there were any new results related to this question. To my amazement I found out that mathematicians had shown the existence of novel superalgebras over fields of characteristic 3 and 5 that have no analogs over the fields of real and complex numbers. Some of these superalgebras can be interpreted as the super analogs of the E-series over the corresponding fields and as simple superconformal algebras beyond six dimensions. I review some of these results in the Appendix \ref{finitechar}.

 \section{ Magnetic  charge and non-associative algebras in quantum mechanics}
      I would like to now turn to discuss my work with Bruno Zumino on the relationship between non-associative algebras and magnetic charges.  It was shown long time ago that the commutators of the velocities of a non-relativistic  electron in the field of a point magnetic monopole  violate  the Jacobi identity at the position where the monopole is located \cite{Lipkin:1969ck}. 

      The commutators of the coordinates $q_a$ and velocities $v_a$ of a charged particle in a magnetic field read as
\eq
[q_a, q_b] =0, \quad  [q_a,v_b]=i\delta_{ab}
\en
and in particular
\eq
[v_a,v_b] =  i \epsilon_{abc} B_c
\en
where $a, b, c = 1,2,3$, and $\vec{B}$ denotes the magnetic field.
The double commutators of the velocities yield the following
 Jacobiator:

\eq
[v_1, [v_2,v_3]] + [v_2, [v_3,v_1]] + [v_3, [v_1,v_2]] = - \vec{\nabla} \cdot \vec{B}
\en
which corresponds to
a non-vanishing three-cocycle  \cite{Grossman:1984fs,Jackiw:1984rd}.

As discussed by Jackiw  \cite{Jackiw:1984rd},
 a gauge-invariant action of the translations is
represented by the operator ($\hbar =1$)
\eq
U(\vec{a}) = \exp(i \vec{a} \cdot \vec{v}).
\en
The translation group generators act on the wave-function
$\Psi(\vec{q})$ of the charged particle ( electron) 
 via
\eq
U(\vec{a}) \Psi(\vec{q}) = \exp(i \vec{a} \cdot \vec{v})
\exp(-i \vec{a} \cdot \vec{p}) \Psi( \vec{q} + \vec{a}).
\en
where  the momentum $\vec{p}$
is $\vec{p} = \vec{v} + \vec{A}$ and $\nabla \times \vec{A} = \vec{B}$.
The three cocycle appears follows from the  successive action of translations
\eq
[U(\vec{a_1})U(\vec{a_2})]U(\vec{a_3}) \Psi(\vec{q})=
\exp(i \alpha_3) U(\vec{a_1})[U(\vec{a_2}))U(\vec{a_3})] \Psi(\vec{q}).
\en
For  a point magnetic monopole,
$\nabla \cdot \vec{B} = 4\pi \delta{(\vec{q})}$,
the three-cocycle is proportional
to the flux out of the tetrahedron formed from the three vectors
$\vec{a_1}, \vec{a_2}, \vec{a_2}$, with one vertex at $\vec{q}$.

Hence the three-cocycle in this case implies 
that the operators  representing  translations become nonassociative as a result of the failure of 
 the Jacobi identity. It was then suggested that one can restore associativity at the global ( integrated) level by requiring the Dirac quantization condition  which leads to 
 $\alpha_3 = 4 \pi e g $ where $g$  is the  magnetic charge of the monopole \cite{Grossman:1984fs,Jackiw:1984rd}.  
 
  However as  was pointed out  by the authors of  \cite{Boulware:1985hp}, 
 if the coordinates and the
 velocities of the electron are represented by linear operators acting on a Hilbert
 space, the Jacobi identity can not be violated. Linear operators acting on a Hilbert space  are
 always associative. Therefore to violate the Jacobi identity 
 operators representing  the coordinates and velocities must  belong to an intrinsically nonassociative algebra of observables.
In my work with Bruno Zumino we investigated this problem to find out what kind of non-associative algebra of observables 
can describe  a non-relativistic electron in the
 field of a magnetic charge distribution
 \cite{Gunaydin:1985ur}. 
 We studied , in particular, the conditions under which   the nonassociative algebraic structure
 defined by velocities and coordinates of an electron in a  magnetic
 charge distribution is  that of a Malcev algebra.
Malcev algebras correspond to generalizations of Lie algebras with an anti-symmetric product $\star$ \cite{malcev,sagle}
\[ a \star b  = - b \star a \]
that satisfy the so-called Malcev identity
\[ (a \star b) \star (a \star c) = ((a \star b) \star c) \star a + ((b \star c) \star a) \star a + ((c \star a) \star a ) \star b \]
which can be rewritten  in the form
\[ J(a,b,a \star c) = J(a,b,c) \star a \]
where $J(a,b,c)$ is the Jacobiator :
\[ J(a,b,c) \equiv ((a\star b)\star c) + (( c \star a) \star b) + (( b \star c) \star a) \]
 For  the electron moving in the field of some magnetic charge distribution the Malcev condition implies
 \[ \vec{\nabla} (\vec{\nabla} \cdot \vec{B}(x)) =0\]
 The case $(\vec{\nabla} \cdot \vec{B}(x))=0$ corresponds to a Lie algebra and the operators can be represented in terms of Hilbert space operators by taking
 \[ v_a = p_a - A_a \qquad B_{ab}= \partial_a A_b -\partial_b A_a \]

 On the other hand the case $ B_a \propto x_a $ i.e $ (\vec{\nabla} \cdot \vec{B}(x))=constant $ leads to a bona fide Malcev algebra. By redefining the coordinates and the velocities one can choose $B_a(x) = x_a$ resulting in what we called the "magnetic algebra":
 \[ [x_a, x_b]=0 \quad , \,  [x_a , v_b]= i \delta_{ab} \]\[
[v_a,v_b]=i \epsilon_{abc} x_c \]
which corresponds to constant magnetic charge distribution. The corresponding quantum mechanical operators can not be represented by operators acting on a Hilbert space
 
\section{Nonassociativity in String Theory and the Magnetic Algebra of G\"unaydin and Zumino }
Constant $B$-field background in string theory leads to a non-commutative but
associative star product
\cite{kont}. Hence the  Seiberg-Witten limit of open string field theory in such backgrounds is described by 
 a noncommutative gauge theory \cite{connes, seibwitten, schom, chu}.
Furthermore it was pointed out in \cite{corn,ho} that the presence of a non-zero $H=dB$ background can give rise to
nonassociative star-product in the singular  limit of vanishing background metric \cite{corn, ho}. Conformal field theories
involving the non-trivial three-form flux corresponding to non-singular consistent string backgrounds have been studied in \cite{lust} and the appearance of nonassociative structures
in such backgrounds have been established.
In a paper I wrote with Djordje Minic \cite{Gunaydin:2013nqa} which we dedicated to Bruno Zumino on the occasion of his 90th birthday in which we pointed out that the non-associative algebra that arises for constant $H$-field in the works of \cite{lust} is isomorphic to the magnetic algebra studied in my paper with Bruno Zumino\cite{Gunaydin:1985ur}, which I shall review briefly below.

The sigma model action describing the open strings propagating in a curved background, that leads to  noncommutative field theories is given by  \cite{connes, seibwitten, schom, chu,corn,ho}\footnote{ With the choice
  ($2\pi \alpha' =1$).}.
\eq
S = {1 \over 2} \int g_{ab}(X) dX^a \wedge *dX^b
+ {i \over 2} \int B_{ab}(X) dX^a \wedge dX^b
\en
where the integration runs over the string worldsheet.
To incorporate a $U(1)$ field at the boundary of the worldsheet one also has to include the term 
\eq
S_b = \int F_{ab}dX^a \wedge dX^b.
\en
 Expanding the metric and the $B$ field around constant backgrounds  in Riemann normal coordinates  to leading order leads to the following action
\eq
S = {1 \over 2} g_{ab} \int dX^a \wedge *dX^b
+ i \int \omega
+{i \over 6} H_{abc} \int X^a  dX^b \wedge dX^c
\en
where the effective symplectic structure  is \cite{corn} 
\eq
\omega_{ab} (x) = B_{ab} + F_{ab} (x)
\en
The 3-form field strength is given by
\eq
H= dB.
\en
which we assume is constant to leading order in weakly curved backgrounds. 

The correlation functions of operators inserted at the boundary with H-flux turned off
are given by 
\eq 
<f_1 f_2...f_n> = \int V(\omega) dx (f_1 * f_2 *
...*f_n), 
\en 
where $V(\omega)$ is the  volume form \cite{corn} and $*$ is the associative Weyl-Moyal-Kontsevitch star product with
respect to the symplectic structure
$\omega^{-1}$
\eq f * g = f g +{i \over 2}
\omega^{ab} \partial_{a}f \partial_{b} g+ ... 
\en 
The associativity is equivalent to the
condition that $\omega$ be  closed, i.e  $d\omega =0$ \cite{corn}. As was shown in \cite{corn} in the limit   $g_{ab} \rightarrow 0$
the  above star product becomes nonassociative.  However the consistent string backgrounds require a non-zero metric  \cite{lust} and in the presence of H-flux  the star product becomes the bullet $\bullet$ product 
\eq
f \bullet g = f g + {i \over 2}
{\tilde{\omega}}^{ab}
\partial_{a}f \partial_{b} g+ ...
\en
where the modified "symplectic metric " ${\tilde{\omega}}_{ab} (x)$ is given by 
\eq
{\tilde{\omega}}_{ab} (x) = B_{ab} + {1 \over 3} H_{abc} x^c + F_{ab} .
\en
for constant  $H$. 
Since ${\tilde{\omega}}$ is not closed, $d {\tilde{\omega}} = H$,
the $\bullet$ product is not associative
\eq
(f \bullet g) \bullet h - f \bullet (g \bullet h) =
{1 \over 6} {\tilde{\omega}}^{ia} {\tilde{\omega}}^{jb} {\tilde{\omega}}^{kc}
H_{abc} \partial_{i} f \partial_{j} g \partial_{k} h + ...
\en

From the world-volume point of view, this implies the  non-commutativity
and nonassociativity of coordinates with the 
 commutator of coordinates given by
\eq
[x^i, x^j]_{\bullet} \equiv x^i \bullet x^j -  x^j \bullet x^i =  i {\tilde{\omega}}^{ij}.
\en
Furthermore, the Jacobi identity is violated for non-zero H-flux
\eq
[x^i, [x^j, x^k]_{\bullet}]_{\bullet} +
[x^j, [x^k, x^i]_{\bullet}]_{\bullet} +
[x^k, [x^i, x^j]_{\bullet}]_{\bullet} = -
{\tilde{\omega}}^{ia}{\tilde{\omega}}^{jb}{\tilde{\omega}}^{kc} H_{abc}.
\en
For  closed string theory compactified on a 3-torus in the constant $H$ background such that
\eq \tilde{\omega}_{ij} \sim H_{ijk} p^k \en where $p$  the momentum \cite{lust}, the resulting non-associative algebra is isomorphic to the G\"unaydin Zumino magnetic algebra \cite{Gunaydin:1985ur}  describing  the motion of
an electron moving  in the field of a constant magnetic charge distribution with the role of coordinates and momenta interchanged.  We should stress again that  the consistency of the conformal field theory involving the constant $H$ field 
crucially relies on the non-trivial gravitational background \cite{lust}.  The more general "magnetic" charge distributions are expected to be described by other Malcev algebras and their generalizations. Furthermore
one can formally generalize the magnetic algebra  in a way that puts the coordinates and momenta on an equal footing\cite{Gunaydin:2013nqa}:
 \[ [x_a, x_b]= - i \epsilon_{abc} E_c \quad , [x_a , v_b]= i \delta_{ab} \quad ,
[v_a,v_b]=i \epsilon_{abc} B_c \]

\[ [[v_1,v_2],v_3]+[[v_3,v_1],v_2]+[[v_2,v_3],v_1]=- \vec{\nabla}_x \cdot \vec{B} \]
\[ [[x_1,x_2],x_3]+[[x_3,x_1],x_2]+[[x_2,x_3],x_1]= \vec{\nabla}_v \cdot \vec{E} \]
which is , no longer, a Malcev algebra. 

We should note that St\"uckelberg gave a generalization of   Poisson brackets in classical statistical mechanics  while preserving the Liouville's theorem such that they no longer satisfy the Jacobi identity. Earlier he proposed a generalization of quantum mechanics  with non-linear operators with a fundamental length. As discussed in \cite{Gunaydin:2013nqa} non-associativity can be considered as a special form of nonlinearity. As such superstring theory combines both ideas of St\"uckelberg. This is briefly reviewed in Appendix B. For further details I refer to \cite{Gunaydin:2013nqa} . 

\section{Uplift of non-associative  R-flux algebra of string theory to M-theory and octonions \label{GLM}}

About a year after the Bruno Zumino memorial meeting at CERN , in a joint work with L\"ust and Malek we showed that the magnetic algebra of G\"unaydin and Zumino can be obtained by contraction from the simple Malcev algebra generated by the seven imaginary units  of octonions\cite{Gunaydin:2016axc}. Since the magnetic algebra is isomorphic to certain non-geometric R-flux algebra that arises in string theory\cite{lust} this raises the question what the physical meaning of the uncontracted Malcev algebra of octonions is. Our study of this question led to the proposal that the uncontracted Malcev algebra of octonions describes the uplift of the string theory R-flux algebra to M-theory flux algebra. Referring to the paper \cite{Gunaydin:2016axc} for details I will review briefly our results  below. 

In our paper we focused mainly on the so-called  parabolic non-geometric string $R$-flux models on a three-torus with constant fluxes. In this case the non-trivial commutation relations among the closed string coordinates and momenta are given by 
\begin{equation}
\left[ x^i, x^j \right] = i \frac{l_s^3}{\hbar} R^{ijk} p_k \,,
\label{commclosed}
\end{equation}
\begin{equation}
\left[x^i, p^j\right]=i \hbar \delta^{ij}\,, \quad \left[p^i, p^j\right]=0 \,.
\label{commclo23}
\end{equation}
where $i_s$ is the string length and $i,j,..=1,2,3$. 
The Jacobiator of the closed string coordinates does not vanish and  is given by:
\begin{equation}
\left[ x^i, x^j, x^k\right] \equiv \frac13 \left[\left[x^1, x^2\right], x^3\right]+{\rm cycl.~perm.} = l_s^3 R^{ijk}\, ,
\label{asso}
\end{equation}

At first sight our proposal that the uplift of the string R-flux algebra to M-theory is described by the uncontracted simple Malcev algebra seems to contradict the expectation that the phase spaces of physical systems are even dimensional while the simple Malcev algebra of octonions is seven dimensional. To support our proposal we  considered a four dimensional toy model for an R-flux background which is dual to a twisted torus and showed that the locally non-geometric background is missing a momentum mode and hence the corresponding phase space is seven dimensional. This makes it possible to identify the four  coordinates and three momenta with the seven imaginary units of octonions in M-theory backgrounds.
Denoting the four coordinates as $X^\alpha$ and three momenta as $P^i$  the full non-associative M-theory flux  algebra in our model takes the form:
\begin{equation}
 \begin{split}
  \left[ P_i, P_j \right] &= - i \lambda \hbar \epsilon_{ijk} P^k \,, \qquad \left[ X^4, P_i \right] = i \lambda^2 \hbar X_i \,, \\
  \left[ X^i, X^j \right] &= \frac{il_s^3}{\hbar} R^{4,ijk4} P_k \,, \qquad \left[ X^4, X^i \right] = \frac{i \lambda l_s^3}{\hbar} R^{4,1234} P^i \,, \\
  \left[ X^i, P_j \right] &= i \hbar \delta^{i}_{j} X^4 + i \lambda \hbar \epsilon^{i}{}_{jk} X^k \,, \\
  \left[ X^i, X^j, X^k \right] &= l_s^3 R^{4,ijk4} X_4 \,, \\
  \left[ X^i, X^j, X^4 \right] &= - \lambda^2 l_s^4 R^{4,ijk4} X_k \,, \\
  \left[ P_i, X^j, X^k \right] &= 2 \lambda l_s^3 R^{4,1234} \delta_{i}^{[j} P^{k]} \,, \\
  \left[ P^i, X^j, X^4 \right] &= \lambda^2 l_s^3 R^{4,ijk4} P_k \,, \\
  \left[ P_i, P_j, X_k \right] &= - \lambda^2 \hbar^2 \epsilon_{ijk} X^4 + 2 \lambda \hbar^2 \delta_{k[i} X_{j]} \,, \\
  \left[ P_i, P_j, X_4 \right] &= \lambda^3 \hbar^2 \epsilon_{ijk} X_k \,, \\
  \left[ P_i, P_j, P_k \right] &= 0 \,. \label{eq:FullMAlgebra}
 \end{split}
\end{equation}
where $l_s$ is the string length , $ R^{4,\alpha\beta\gamma\delta}=N \epsilon^{\alpha\beta\gamma\delta}$ is the M-theory $R$-flux tensor ( $\alpha , \beta , ..=1,2,3,4$ ) and $N$ is a parameter such that the four coordinates $X^\alpha$ and three momenta $P^i$ are related to the 7 imaginary units $( e_i , e_{i+3} , e_7 ) $ of octonions as
\begin{equation}
 X^i = \frac12 i \sqrt{N} e_{i+3} \,, \qquad X^4 =  \frac12 i \sqrt{N} e_7 \,, \qquad P_i = - \frac12 i e_i \,, \label{eq:XPnodimensions}
\end{equation}
where $i = 1, \ldots, 3$.
In the limit $\lambda \rightarrow 0$  this M-theory  R-flux algebra reduces to the string $R$-flux algebra. The $\lambda \rightarrow 0$  corresponds to the string coupling $g_s$  going to zero limit.  We conjectured that $\lambda$ remains finite as $g_s \rightarrow \infty$ so that the Malcev algebra description remains relevant  in the strong-coupling regime as well.
Later more examples of non-geometric M-theory backgrounds with missing momenta were studied in \cite{Lust:2017bwq}.

\section{ Concluding Remarks}
As we discussed above the exceptional Jordan algebra over the division algebra of octonions determines uniquely the exceptional supergravity and  the exceptional Jordan algebra over the split octonions underlies the maximal supergravity and describes the  internal space-time, in the sense of \cite{Gunaydin:1975mp}, of the  $E_{6(6)}$ covariant  extension of eleven dimensional supergravity theory  in the double field theory formulation\cite{Hohm:2013vpa}. 

Octonionic quantum mechanics defined over the exceptional Jordan algebra describes two octonionic degrees of freedom corresponding to the Moufang plane\cite{Gunaydin:1978jq} and is unique in the sense that it can not be embedded in a higher dimensional quantum mechanics since infinite dimensional extension of the exceptional Jordan
algebra does not exist. This is also reflected in the fact that Moufang plane describes a non-Desarguesian projective geometry and hence can not be embedded in a higher projective geometry unless one modifies the axioms of quantum mechanics so that they are no longer equivalent to axioms of projective geometry.  

As pointed out in \cite{Gunaydin:2013nqa} and reviewed above the superstring theory that admits non-associative flux backgrounds with a fundamental length might be interpreted as a realization  of St\"uckelberg's proposal for a non-linear extension of quantum mechanics with a fundamental length if one interprets non-associativity as a very special form of non-linearity. The fact that these backgrounds admit extensions to M-theory as was shown in \cite{Gunaydin:2016axc} suggest further that the axiomatic formulation of quantum mechanics underlying M/superstring theory would require modifications of the axioms of quantum mechanics as formulated by Birkhoff and von Neumann\cite{MR1435976}.

The deep mathematical connections between supersymmetry, exceptional groups and non-associative algebras might be a hint that the  relevance of exceptional groups and related algebraic structures in describing Nature at a fundamental level is  intimately tied to the relevance of supersymmetry in describing Nature. 

We may not know the answer to these questions for a long time to come. However, intellectually  it has been most rewarding for me to work on supersymmetric theories and contribute to uncovering some of the remarkable connections  between  geometry, exceptional groups, representation theory  and supersymmetry.




\section{ Appendices} 
\appendix

\section{\label{finitechar} Superconformal symmetry beyond six dimensions and super extensions of the  Magic Square over finite fields}
Benkart and Zelmanov showed that Tits construction of the Magic Square\cite{Tits66}  can be extended by replacing the degree three Jordan algebras over associative composition algebras with Jordan superalgebras $J^{(1|2)}$ and $ JD(2/2)_\alpha$ where $J^{(1|2)}$ is the Heisenberg algebra, if the characteristic of the underlying field $\mathbb{F}$ is not equal to 2,3 or 5\cite{BZ96}. Later Elduque showed that over a field of characteristic 5 the Tits construction extends also to the exceptional Jordan superalgebra $JF(6/4)$\cite{Eld07b}. 
Hence over fields of characteristic $\neq 2,3$ we have a super Magic Rectangle give in Table \ref{magicrectangle} that extends the Magic Square which we reproduce from reference \cite{Elduque_Tits}.

  \begin{table}
  \begin{center}
  \normalsize
  \begin{center}
  {\bf \color{red} Super Magic Rectangle }
  \end{center}
  \begin{displaymath}
\begin{array}{|c|c|c|c|c|c|c|c|}
\hline
~ & J_3^{\mathbb{F}} & J_3^{\mathbb{F\times F} } &J_3^{\mathbb{M(F)}_2} &J_3^{\mathbb{O(F)}}& JD(1|2) & JD(2|2)_\alpha & JF(6|4)  \\ \hline
&&&&&&& \\
\mathbb{F} & SO(3) & SU(3)& USp(6)  & F(4)&Sp(2)&OSp(1|2)& OSp(1|2)^2\\
\hline
&&&&&&& \\
 \mathbb{F\times F} & SU(3)& SU(3)\times SU(3)  & SU(6) & E_6 &OSp(1|2) &SU(2|1) &OSp(2|4) \\
 \hline
 &&&&&&& \\
  \mathbb{M(F)}_2 & USp(6)& SU(6) & SO(12) &  E_7&OSp(3|2) &D(2,1;\alpha)& F(4)\\
 \hline
 &&&&   &&&  \\
  \mathbb{O(F)} &F(4)& E_6 & E_7 & E_8 &G(3)&F(4)_{\alpha=2}&\color{red} \mathcal{T}(55|32) (ch=5) \\
 \hline
\end{array}
\end{displaymath}
  \end{center}
\caption{\label{magicrectangle} The Magic Rectangle is an extension of Magic Square by replacing the Jordan algebras of degree three in Tits construction with Jordan superalgebras $ J^{0|2} , JD(2|2)_\alpha $ and  $ JF(6|4)$. The last column exist only over a field of characteristic 5.$\mathbb{F}$ denotes the underlying field and $\mathbb{M}(\mathbb{F})_2$ denotes the "quaternion algebra"  over $\mathbb{F})$}
\end{table}

The label $\mathcal{T}(55|32)_5$ stands for a simple Lie superalgebra whose even subalgebra is $SO(11)$  and whose odd elements are in the spinor $32$ representation of $SO(11)$  in characteristic five\cite{Elduque_2007,Elduque_Tits,Cunha_Elduque}. 
It can be considered as an  AdS superalgebra in ten dimensions or conformal superalgebra  in nine dimensions with the  odd generators in the spinor representation $32$ in characteristic five. It can also be interpreted as the quasiconformal superalgebra associated with the unique exceptional Jordan superalgebra $JD(6/4)$ just like $E_8$ is the quasiconformal Lie algebra associated with the exceptional Jordan algebra $J_3^{\mathbb{O}}$.

There is another  extension of the Magic Square to an extended super Magic Square over  fields of characteristic 3. What makes characteristic 3 special is the fact that nontrivial composition superalgebras exist only over fields of characteristic three. More specifically there exist two such composition superalgebras of dimension 3 and and 6 denoted as $B(1,2)$ and $B(4,2)$, respectively.  The Kantor construction of the Magic Square over the tensor product of two composition algebras were shown to extend extend to the super Magic Square by including the two composition superalgebras which we reproduce in Table \ref{supermagicsquare} following \cite{Elduque_Tits}. For further references on the extensions of Magic square over finite fields see the papers \cite{Elduque_2007,Elduque_Tits,Cunha_Elduque}.

  \begin{table}
  \normalsize
  \begin{center}
  {\bf \color{red} Super Extended  Magic Square}
  \end{center}

  \begin{displaymath}
\begin{array}{|c|c|c|c|c|c|c|}
\hline
~ & \mathbb{F} & \mathbb{F\times F}  &\mathbb{M(F)}_2 &\mathbb{O(F)}& B(1,2) & B(4,2) \\ \hline
&&&&&& \\
\mathbb{F} & SO(3) & SU(3)& USp(6)  & F(4)& psl_{2,2} &sp_6\oplus (14)\\
\hline
&&&&&& \\
 \mathbb{F\times F} & SU(3)& SU(3)\otimes SU(3)  & SU(6) & E_6 & (pgl_3 \oplus sl_2) \oplus (psl_3 \otimes (2)) &pgl_6 \oplus (20)   \\
 \hline
 &&&&&& \\
 \mathbb{M(F)}_2 & USp(6)& SU(6) & SO(12) &  E_7&(sp_6 \oplus sl_2) \oplus ((13)\otimes (2)) &\color{red} so(12)\oplus (spin_{12})\\
 \hline
 &&&&   &&  \\
  \mathbb{O(F)} &F(4)& E_6 & E_7 & E_8 & (f_4 \oplus sl_2 ) \oplus ( (25)\otimes (2)) &\color{red}  e_7 \oplus (56)  \\   \hline
  &&&&&& \\
  B(1,2) &&&&& so(7) \oplus 2 (spin_7) & sp_8 \oplus (40) \\ \hline
  &&&&&& \\
  B(4,2) &&&&& & \color{red} so(13) \oplus spin_{13} \\
 \hline
\end{array}
\end{displaymath} 
\caption{\label{supermagicsquare}  Super extended Magic Square that extend the Kantor construction of the Magic Square over the tensor product of two composition algebras by including composition superalgebras $B(1|2)$ and $B(4|2)$ that exist only over fields of characteristic three. Note that $spin_d$ denote the spinor representation of $SO(d)$ over the corresponding field.  The notation $g \oplus (n)  $ indicates a simple superalgebra with even subalgebra $g$ and odd elements transforming in $n$ dimensional representation of $g$. The super magic square above is symmetrical and some entries were left out. }
\end{table}

 Superalgebra $SO(10,2) \oplus (32) $ in Table \ref{supermagicsquare} can be interpreted as $AdS$ superalgebra in 11 dimensions or as conformal superalgebra in ten dimensions with 32 supercharges. Similarly the superalgebra $SO(13) \oplus spin_{13} $ can be interpreted as $AdS_{12}/Conf_{11}$ superalgebra with 64 supercharges. 
The simple superalgebra  $ E_7 \oplus (56)$ decribes a conformal superalgebra over the exceptional space-time with 56 supercharges over a field of characteristic 3. Note that 
$SO(12) \oplus (32) \subset E_{7} \oplus 56 $. None of the simple  $AdS_{d+1}/CFT_d$ superalgebras for $d\geq 9$ that exist over fields of characteristic 3 or five have counterparts over fields of characteristic zero which include the reals $\mathbb{R}$ and complex numbers $\mathbb{C}$. Whether or how these novel mathematical structures could be relevant to describing physical phenomena is an open problem. 
\section{ St\"uckelberg's generalized poisson brackets and string theory}
Long time ago St\"uckelberg proposed a  generalization of  Poisson brackets (PB) in  classical statistical mechanics while preserving Liouville's theorem such that the PB's  no longer satisfy the Jacobi identity \cite{ecgs}. Earlier St\"uckelberg also considered a possible nonlinear extension of quantum mechanics with a fundamental length $\lambda_0$  such that uncertainties in the measurements of coordinates satisfy  \cite{stuckelberg2} 
\begin{equation}
(\Delta X)^2 \eqslantgtr  ( \lambda_0 )^2 \nn
\end{equation}
and proposed modifying the minimum uncertainty relation in one dimension as:
\footnote{In three dimensions he proposes  using the minimum uncertainty among all possible directions.}
\begin{equation}
(\Delta X)^2 (\Delta P)^2 = \frac{\hbar^2}{4} \left( 1 - \frac{(\lambda_0)^2}{(\Delta X)^2} \right)^{-1} \nn
\end{equation}
which requires the modification of canonical commutation relations as
\begin{equation}
i [ P, X] = \hbar  \left( 1 - \frac{(\lambda_0)^2}{(\Delta
X)^2} \right)^{-1/2}. \nn
\end{equation}
If one expand the inverse $ \left( 1 - \frac{(\lambda_0)^2}{(\Delta X)^2} \right)^{-1}$
and use to first order $\Delta X \sim \hbar \, \Delta P^{-1}$ we  get the stringy
uncertainty relation which usually reads as\cite{Gunaydin:2013nqa}
\begin{equation}
\Delta X \Delta P \sim (1+ \alpha' \Delta P^2) \hbar. \nn
\end{equation}
(This formal procedure relates $\lambda_0$ with the string scale $ l_s$,
or equivalently with $\alpha' \sim l_s^2$.)
As pointed out in \cite{Gunaydin:2013nqa} non-associativity can be considered as a particular form of nonlinearity and , hence.  if one replaces non-linearity with non-associativity then string theory with a fundamental length corresponds, in a very broad sense,  to the kind of quantum theory St\"uckelberg was envisioning.

\newpage



\def\cprime{$'$} \def\cprime{$'$} \def\cprime{$'$} \def\cprime{$'$}
\providecommand{\href}[2]{#2}\begingroup\raggedright\endgroup

\end{document}